\documentclass[reprint,amsmath,amssymb,aps,superscriptaddress,twocolumn,nofootinbib,10pt]{revtex4-1}

\usepackage[colorlinks=true,allcolors=blue]{hyperref}
\usepackage{enumitem}
\usepackage{graphicx,color}
\usepackage{dcolumn}
\usepackage{subfigure}
\usepackage{bm}
\usepackage{amsmath,amsfonts,amssymb}
\usepackage{acronym}
\usepackage{enumitem}
\usepackage{array}
\usepackage{multirow}
\usepackage{CJK}
\usepackage{array}
\usepackage{orcidlink}
\usepackage{xcolor}
\usepackage{booktabs}
\newlist{alphalist}{enumerate}{1}
\setlist[alphalist,1]{label=\textbf{\alph*.}}
\allowdisplaybreaks
\newcommand{\be}{\begin{equation}}
\newcommand{\ee}{\end{equation}}
\newcommand{\bea}{\begin{eqnarray}}
\newcommand{\eea}{\end{eqnarray}}

\newcommand{\SOUTHCUT}{School of Physics and Optoelectronics, South China University of Technology, Guangzhou 510641, People's Republic of China}
\newcommand{\NCUa}{Department of physics, Nanchang University, Nanchang, 330031, China}
\newcommand{\NCUb}{Center for Relativistic Astrophysics and High Energy Physics, Nanchang University, Nanchang, 330031, China}
\newcommand{\IIT}{Indian Institute of Technology, Gandhinagar, Gujarat-382355, India}

\newacro{EMRI}{extreme mass-ratio inspirals}
\newacro{MBH}{massive black hole}
\newacro{BH}{black hole}
\newacro{GR}{general relativity}
\newacro{HKBH}{hairy Kerr black hole}
\newacro{KNBH}{Kerr-Newmann black hole}
\newacro{KBH}{Kerr black hole}
\newacro{NHT}{no-hair theorem}
\newacro{DWD}{double white dwarf}
\newacro{GW}{gravitational wave}
\newacro{AK}{analytic kludge}
\newacro{NK}{numerical kludge}
\newacro{AAK}{augmented analytic kludge}
\newacro{CO}{compact object}
\newacro{PE}{parameter estimation}
\newacro{SNR}{signal-to-noise ratio}
\newacro{PN}{post newtonion}
\newacro{FIM}{Fisher information matrix}
\newacro{LSO}{last stable orbit}
\newacro{ISCO}{innermost stable circular orbit}
\newacro{BBH}{Binary Black Hole}
\newacro{BNS}{Binary Neutron Star}
\newacro{NS}{Neutron Star}
\newacro{KN}{Kerr-Newmann}
\newcommand{\beq}{\begin{equation}}
\newcommand{\eeq}{\end{equation}}
\newcommand{\beqa}{\begin{eqnarray}}
\newcommand{\eeqa}{\end{eqnarray}}

\def\lsim{\mathrel{\rlap{\lower4pt\hbox{\hskip0.5pt$\sim$}}
    \raise1pt\hbox{$<$}}}         
\def\gsim{\mathrel{\rlap{\lower4pt\hbox{\hskip0.5pt$\sim$}}
    \raise1pt\hbox{$>$}}}         
\begin{document}
\begin{CJK*}{UTF8}{gbsn}
\title{{\Large {\bf
Extreme mass-ratio inspirals and extra dimensions: Insights from modified Teukolsky framework
}}}

\author{Shailesh Kumar\orcidlink{0000-0001-7072-9452}}
\email{shailesh.k@iitgn.ac.in}
\affiliation{\IIT}

\author{Tieguang Zi\orcidlink{0000-0003-0046-2056}}
\email{zitieguang@ncu.edu.cn}
\affiliation{\NCUa}
\affiliation{\NCUb}
\affiliation{\SOUTHCUT}

\author{Arpan Bhattacharyya\orcidlink{0000-0002-7933-6441}}
\email{abhattacharyya@iitgn.ac.in}
\affiliation{\IIT}

\begin{abstract}
Extreme mass-ratio inspirals (EMRIs) offer a promising avenue to test extra-dimensional physics through gravitational wave (GW) observations. In this work, we study equatorial eccentric EMRIs around a spherically symmetric braneworld black hole, focusing on the influence of a tidal charge parameter arising from extra dimensions. Using the fact of $tr$-symmetry of the spacetime under consideration, we implement the Modified Teukolsky Equation (MTE) framework, incorporating the non-Ricci-flat nature of the spacetime. We compute the relevant observables and perform a comparative analysis with the results obtained from the Dudley-Finley (DF) approximation. Our findings indicate that the constraint on the tidal charge remains nearly the same in both approaches\textemdash MTE and DF\textemdash thus supporting previous studies on EMRIs in braneworld scenarios within the DF approximation. Furthermore, the difference in the mismatch between the two formulations exhibits deviations as the orbital eccentricity increases. Therefore, these findings highlight not only the observational potential of future low-frequency detectors like the Laser Interferometer Space Antenna (LISA) but also bring out the effectiveness of the DF approximation as well as the importance of the MTE framework for accurately modeling binaries in theories beyond GR.
\end{abstract}
\maketitle
\end{CJK*}
\section{Introduction}
The ongoing advancements in determining distinct characteristics of black holes through gravitational wave (GW) observations have opened several avenues for testing General Relativity (GR) and, beyond that, offer deep insights into the behaviour of compact objects in the strong gravity regime. GW detections have not only tested predictions of GR but also enabled us to explore the nature and population of binary black hole systems \cite{LIGOScientific:2016aoc,Uchikata:2019frs, LIGOScientific:2016lio, LIGOScientific:2020tif, LIGOScientific:2021sio, LIGOScientific:2019fpa, LIGOScientific:2016sjg, LIGOScientific:2017bnn, LIGOScientific:2018jsj}. Although many such events have already been observed, a promising class of sources\textemdash known as extreme-mass-ratio inspirals (EMRIs)\textemdash has emerged with growing interest due to their unique physical characteristics and observational potential \cite{PhysRevD.105.L061501, PhysRevLett.129.241103, Destounis:2022obl, Rahman:2023sof, Duque:2024mfw, Dai:2023cft, Miller:2025yyx, Cardenas-Avendano:2024mqp, PhysRevD.102.064041, PhysRevLett.126.141102, Kumar:2024utz, Zhang:2024csc, Kumar:2024our, Zi:2023qfk, Zi:2024jla, Kumar:2024dql, Zi:2022hcc, Zi:2023pvl, Fu:2024cfk, Qiao:2024gfb, Babichev:2024hjf}. The mass-ratio of such binaries lies in the range $q=  (10^{-7} - 10^{-4})$. The characteristic low-frequency signals from these sources align well with the sensitivity range of space-based detectors, making observatories like the Laser Interferometer Space Antenna (LISA) \cite{LISA:2017pwj} and TianQin \cite{TianQin:2015yph, TianQin:2020hid} strong contenders for their potential detection in the near future. The inspiral phase is one of the crucial stages in the evolution of an EMRI system, during which the orbit shrinks slowly over time while sweeping through a large number of cycles. This phase not only contributes the bulk of the GW signal but also allows accurate extraction of physical parameters like the masses, spins, and orbital eccentricity of the system \cite{Sathyaprakash:2009xs, Bailes:2021tot}. Theoretical modeling of this regime often employs perturbative approaches, such as black hole perturbation theory \cite{Pound:2021qin, Blanchet:2013haa}, which treats the small object as a perturbation in the background of the larger black hole. In particular, the Teukolsky formalism provides an efficient and powerful method to compute GW fluxes in such setups, especially in axisymmetric and stationary spacetimes \cite{Teukolsky:1973ha, 1972ApJ, Teukolsky:1972my, Press:1973zz, Teukolsky:1974yv}.

We know that the original/standard Teukolsky formalism was developed specifically within the framework of GR, applicable to background spacetimes satisfying the vacuum Einstein equations and belonging to the Petrov type D and Ricci flat class--such as the Schwarzschild/Kerr geometry, where for principal null direction aligned tetrads, all Weyl scalars vanish except $\Psi_{2}$ \cite{Teukolsky:1973ha, 1972ApJ, Teukolsky:1972my, Press:1973zz, Teukolsky:1974yv}. However, while examining dynamics in non-GR gravity theories, generally these conditions no longer hold. In such cases, the field equations typically include terms that alter the spacetime geometry, resulting in deviations from Einstein's vacuum equations. Consequently, the black hole solutions may not be Petrov type D and can exhibit more general algebraic classifications, such as Petrov type I. Even if the spacetime is Petrov type D, it can be non-Ricci-flat, contrary to GR black hole solutions. In these spacetimes, vacuum Ricci-flat refers to a configuration where the energy-momentum tensor vanishes and the Ricci tensor is zero, $R_{\mu\nu}=0$, whereas a non-Ricci-flat spacetime has $R_{\mu\nu}\neq 0$, for example due to bulk gravitational effects in braneworld scenarios even in the absence of brane matter. This distinction will be important for the present study because we aim to investigate the effect of this non-Ricci flatness of the background metric on the perturbation equations and, consequently, the gravitational waveforms.

The standard Teukolsky equation fails to capture the full behavior of perturbations, as it does not account for the additional curvature components or modifications to the gravitational interaction. So one requires a generic method/framework, equivalent to the standard Teukolsky equation, that can deal with theories beyond GR. In this line of endeavor, the authors in \cite{Li:2022pcy, Hussain:2022ins} provided the ``modified Teukolsky framework'' to handle non-Ricci-flat Petrov type I and non-Ricci-flat Petrov type D black holes where the geometry can be considered as a linear perturbation of a Petrov type D spacetime in GR. A similar strategy has been used in the context of Hawking radiation for spherically symmetric spacetimes \cite{Arbey:2021jif}. Additionally, the Teukolsky framework, in general, modified Teukolsky equation (MTE), offers several advantages over traditional metric perturbation methods, particularly in the context of GW studies. Unlike metric perturbations, it provides a gauge-invariant, compact description of radiative degrees of freedom through curvature scalars like $\Psi_{4}$ and is written as a scalar equation using the Newman-Penrose (NP) tetrad \cite{Newman:1961qr, Teukolsky:1973ha, Teukolsky:1972my}, where $\Psi_{4}$ is the NP scalar that characterizes outgoing gravitational radiation at null infinity. This makes it especially well-suited for studying quasinormal modes (QNMs) and GW emission in GR and beyond. Recent investigations of QNMs demonstrate that the MTE offers a robust and reliable approach for studying perturbations in modified theories of gravity \cite{Li:2025fci, Wagle:2023fwl}. Therefore, the MTE framework would enable the systematic study of perturbations and GW signatures in a wide class of alternative theories.

In parallel, the progress in high-energy theoretical physics suggests the possibility that our four-dimensional universe is embedded in a higher-dimensional spacetime \cite{Kaluza:1921tu, Klein1926}. Such motivation, stemming from the string theory and M-theory frameworks, has given rise to braneworld models, where all Standard Model fields are confined to a 4-dimensional hypersurface (the brane), while gravity propagates through the full higher-dimensional bulk \cite{Shiromizu:1999wj, Rubakov:2001kp}. In this scenario, the effective gravitational field equations on the brane contain an additional term, the projected electric part of the bulk Weyl tensor ($E_{\mu\nu}$) that physically carries nonlocal information about the bulk gravitational field and represents how the five-dimensional geometry influences the dynamics on the four-dimensional brane. This tensor is symmetric, traceless, and divergence-free \cite{Shiromizu:1999wj, Dadhich:2000am}. An interesting feature of such solutions is the emergence of a parameter known as the ``tidal charge,'' ($Q$) captured in $E_{\mu\nu}$, which arises due to the influence of the bulk geometry and serves as a potential signature of extra dimensions \cite{Dadhich:2000am, Chamblin:2000ra, Aliev:2005bi, Harko:2004ui}. This solution closely resembles the Reissner-Nordstr$\Ddot{\textup{o}}$m (RN) metric, differing only by the presence of a tidal charge term that effectively replaces the electric charge and can take either sign \cite{Dadhich:2000am}.

This work is motivated by the possibility of using EMRIs as precision tools to search for non-GR effects, an extension of previously existing works \cite{Rahman:2022fay, Zi:2024dpi, Kumar:2025njz}. As per the infrastructure available in \cite{Rahman:2022fay, Zi:2024dpi, Kumar:2025njz}, we know that the effective field equations on the brane include the projected Weyl term $E_{\mu\nu}$, so the geometry is generally not Ricci-flat and the standard Teukolsky equation cannot be applied outright. Although it is nontrivial to solve the perturbation equations in their full generality, we can make some progress by making some well-motivated approximations. For perturbations (in vacuum $\delta G_{\mu\nu}+\delta E_{\mu\nu}=0$), we can assume the perturbed Weyl tensor $\delta E_{\mu\nu}=0\,.$ This assumptions can be motivated in the following way: from the point of view of the higher-dimensional bulk, $\delta E_{\mu\nu} \approx \left(\tfrac{l}{L}\right) G_{\mu\nu} \ll G_{\mu\nu}$, where $l$ is the curvature length scale of the bulk spacetime, $L$ is the curvature length scale associated with the supermassive black hole on the brane and $G_{\mu\nu}$ is the Einstein tensor \cite{Kanno:2002ia, Kanno:2003au}. This approximation is reasonable in the low-energy limit (as well as phenomenologically motivated \cite{Kanno:2002ia, Kanno:2003au}), enabling us to ignore the variation of $\delta E_{\mu\nu}\,.$ {\it Note that the above-mentioned length scale arguments translate into the statement of the smallness of the tidal charge $Q$ as the Weyl tensor for our braneworld black hole is proportional to $Q$, i.e, $\delta E_{\mu\nu}\sim Q \delta\hat{E}_{\mu\nu}$, where, in this paper, we have taken only the leading-order contribution of $Q$ from the perturbation $\delta\hat{E}_{\mu\nu}$} \cite{Dadhich:2000am}. We further emphasize that, in the present analysis, we are working with the 4D effective braneworld geometry where the tidal charge $Q$ carries the imprints of extra dimensions which we keep perturbatively small, which we will also elaborate it further at the later stage of the article. The mentioned assumption is also equivalently understood through the ``Dudley-Finley'' (DF) approximation, where Dudley and Finley originally studied perturbations under the approximation of fixing either the geometry or additional fields such as electromagnetic \cite{Dudley:1977zz, Dudley:1978vd},
consequently, keeping the braneworld tensor field fixed. Hence, under this approximation, we can ignore the variation of the $E_{\mu\nu}$ (that is $\delta E_{\mu\nu}=0$) as a first approximation. Thus, under this approximation, the equation reduces to $\delta G_{\mu\nu}=0$, allowing a Teukolsky-like equation to be used on a Schwarzschild/Kerr-like background with modifications entering only through $\Delta(r)$. We call this the ``DF-Teukolsky''  equation as it retains the same mathematical structure as the Schwarzschild/Kerr Teukolsky case, but with a deformed metric function ($\Delta(r)$) that depends on the braneworld parameter $Q$. We reserve the terminology ``GR Teukolsky'' for the standard Ricci-flat vacuum GR case with $Q=0$, i.e., Schwarzschild/Kerr. At this point, note that from the point of view of the effective 4-dimensional black hole metric (rotating as well as non-rotating) on the brane, the tidal charge parameter $Q$ captures the effect of beyond GR  physics and whatever additional contribution may come if one considers the variation of $\delta E_{\mu\nu}$ will be proportional to it. Hence, as long as $Q$ is very small, the DF approximation can be thought of as the leading order approximation and turns out to be quite effective. We will provide a comparative analysis on this shortly.

Additionally, previous studies on quasinormal modes within the DF approximation for the charged black holes (both rotating and non-rotating) \cite{Berti:2005eb, Saha:2025nsg} showed that small values of the charge parameter yield QNMs that are numerically consistent with those obtained via the metric perturbation method. Authors in \cite{Berti:2005eb, Saha:2025nsg} indicate that the Teukolsky-type equation formulated by Dudley and Finley provides a reliable approximation for the behavior of rotating charged black holes as long as the charge is $\lesssim M/2$. Authors also claim that the DF approximation should be adequate for astrophysical applications. Furthermore, the authors of \cite{Mishra:2021waw} investigate QNMs for rotating braneworld black holes within the DF approximation by considering larger tidal charge values. However, in our present analysis, we consider even smaller values of the tidal charge (several orders of magnitude smaller) in view of constraining it through the LISA. This observation further supports our earlier investigations based on the DF approximation \cite{Rahman:2022fay, Zi:2024dpi, Kumar:2025njz}.

On the other hand, this work has a distinct motivation where we aim to take one further step forward and relax the above-mentioned assumption, using the MTE framework to incorporate the effect of the non-Ricci flatness induced by braneworld corrections (tidal charge $Q$) and investigate its possible effects on the gravitational waveform. \cite{Li:2022pcy, Cano:2023tmv, Guo:2024bqe, Guo:2023wtx}. That is, we relax the assumption of the DF approximation in the present work. To the best of our knowledge, the present work is a first step toward exploring EMRIs in a non-GR background that maintains spherical $tr$-symmetry \cite{Arbey:2021jif} using the MTE framework. The $tr$-symmetry of the background metric allows us to handle the geometry more tractably within the MTE framework, where the product of the temporal and radial metric components becomes $g_{tt}g_{rr}=-1$. This formulation retains the advantages of the original Teukolsky formalism, such as separability \cite{Guo:2024bqe, Guo:2023wtx}, while embedding the modified metric structure arising from the tidal charge parameter. It is worth mentioning that we only include the effect of the non-Ricci-flatness of the spacetime (which it inherits from the background $E_{\mu\nu}$ projected on the brane) in the MTE. To be more precise, utilizing the $tr$-symmetry, one explicitly incorporates the contribution of $\Phi_{11}$ in the MTE, which encodes the information of non-Ricci-flatness of the braneworld geometry. Note that $\Phi_{11}$ is the Ricci scalar that depends on the NP tetrad (it will be discussed in more detail in Section (\ref{app:MTE})). After that, using this MTE, we compute the GW fluxes and waveform characteristics for eccentric equatorial inspirals and systematically compare our results with those obtained from the DF-Teukolsky equation. This comparison not only helps us to quantify the imprint of the tidal charge on observables, but also clarifies the extent to which future detectors like LISA can distinguish such deviations from GR, besides helping us to understand the effectiveness of the DF approximation, as well as underlining the importance of employing the MTE for more precise computation of GW observables for compact binary systems in a theory beyond GR.

However, a further analysis should incorporate full information of $\delta E_{\mu\nu}$, carrying information beyond non-Ricci-flatness, such as capturing the modifications to the Einstein-Hilbert action \cite{Weller:2024qvo} and related aspects. Therefore, obtaining an MTE in the braneworld context remains technically challenging, primarily due to the lack of separability of the perturbation equations. Systematic inclusion of the full contribution coming from $\delta E_{\mu\nu}$, in this \textit{perturbative} framework, as developed in \cite{Li:2022pcy, Wagle:2023fwl}, based on two parameter expansion scheme, might provide us with tools for a more precise computation of the GW observables. While such an analysis lies beyond the scope of the present work, it represents an important direction for future work.

\par
\subsection*{Overview: existing vs present investigation}
Before proceeding further, we would like to provide an overview of this paper as well as make a comparison with our previous work \cite{Rahman:2022fay, Zi:2024dpi, Kumar:2025njz}. We perform the modified Teukolsky analysis for the spherically symmetric braneworld black hole, exploiting its $tr$-symmetry. We will also make a {\it comparison} with our previous results obtained using the DF-Teukolsky equation, whose mathematical form coincides with that of the Teukolsky equation, with the tidal charge entering through the function $\Delta(r, Q)$. The present work advances previous EMRI analyses by {\it relaxing} this approximation and incorporating the non-Ricci-flat character of the {\it  non-rotating} braneworld geometry. A direct comparison between rotating braneworld black holes with DF approximation ($\delta E_{\mu\nu}=0$) and $\delta E_{\mu\nu}\neq0$ remains an open problem, as spin ($a$) breaks $tr$-symmetry and constructing a modified Teukolsky framework for such non-$tr$-symmetric or rotating spacetimes beyond GR is still a significant theoretical challenge, although the DF approximation remains effective for such case as we are probing a very small value of $Q$ \cite{Berti:2005eb} as emphasized earlier. \vspace{0.2cm}\\
We further summarize and restate the recent progress on EMRIs involving braneworld black holes in Table~(\ref{tab1}), also outlined in the following points to clarify the context of the present paper:
\renewcommand{\arraystretch}{1.2}
\begin{table*}[t]
	\centering
    \caption{EMRIs in braneworld Spacetime with Equatorial Orbits \footnote{For all cases, the constraint (i.e how small can the tidal charge be such that it can be detected by LISA) on the tidal charge ($Q$) depends on the black hole spin ($a$), eccentricity ($e$), and the mass-ratio as well as the given detection threshold for mismatch. The related information can be found in \cite{Rahman:2022fay, Zi:2024dpi, Kumar:2025njz}. The mismatch consistently remains larger for the MTE case than the DF-Teukolsky (can be seen in plots in Section (\ref{wave})). Note that the considered values of $Q$ are pretty small in view of their detectability from LISA observations; hence, Dudley-Finley approximation holds well \cite{Berti:2005eb, Saha:2025nsg}.}}
	\begin{tabular}{|c|c|c|c|}
		\hline
		\textbf{Framework} & \textbf{Equatorial Orbits} & \textbf{Spherically Symmetric} & \textbf{Rotating} \\
		\hline
		\multirow{5}{*}{\begin{tabular}{c}
				{\bf DF-Teukolsky} \\
				($\delta E_{\mu\nu}=0$)
		\end{tabular}}
		&   \rule{0pt}{14pt}  Circular orbits
		& \(\checkmark\) \cite{Rahman:2022fay}
		& \(\checkmark\) \cite{Zi:2024dpi} \\
		&
		& Constraint on \(Q \sim 10^{-6}\)
		& Constraint on \(Q \sim 5\times10^{-7}\) \\
		\cline{2-4}
		&  \rule{0pt}{17pt}  Eccentric orbits
		& \(\checkmark\) [{\color{purple}Present Manuscript}]
		& \(\checkmark\) \cite{Kumar:2025njz} \\
		&
		& \(e_0 = 0.1: Q \sim 10^{-5}\)
		& \((a = 0.1, e_0 = 0.4): Q \sim 10^{-5}\) \\
		&
		& \(e_0 = 0.85: Q \sim 5\times10^{-6}\)
		& \((a = 0.9, e_0 = 0.4): Q \sim 10^{-6}\) \\
		\hline
		
		\multirow{4}{*}{\begin{tabular}{c}
				{\bf Modified Teukolsky} \\
				($\delta E_{\mu\nu}\neq 0$)
		\end{tabular}}
		& \rule{0pt}{17pt}  Circular orbits
		& \rule{0pt}{17pt} Follows from Eccentric case
		& \rule{0pt}{17pt} \(\times\) \\
		\cline{2-4}
		& \rule{0pt}{17pt}  Eccentric orbits
		& \(\checkmark\) [{\color{purple}Present Manuscript}]
		&  \\
		&
		& \(e_0 = 0.1: Q \sim 10^{-5}\)
		&  \\
		&
		& \(e_0 = 0.85: Q \sim 5\times10^{-6}\)
		& \(\times\) \\
		\hline
	\end{tabular}\label{tab1}
\end{table*}
\begin{itemize}
    \item Within the Dudley-Finley approximation ($\delta E_{\mu\nu}=0$), circular equatorial orbits for both spherically symmetric and rotating braneworld black holes have been investigated in \cite{Rahman:2022fay, Zi:2024dpi}. The \textit{eccentric equatorial case} within the same approximation has also been examined in \cite{Kumar:2025njz}. Note that the approximation applies at the perturbation level ($\delta E_{\mu\nu}=0$) and the tidal charge contributes from the effective metric function $\Delta (r, Q)$ as the background spacetime still remains the solution of $G_{\mu\nu}+E_{\mu\nu}=0$.

    \item In the present article, using the MTE, which relaxes the DF approximation by allowing $\delta E_{\mu\nu}\neq 0$ (incorporating the contribution of non-Ricci-flatness), we study ``eccentric equatorial orbits'' in the spherically symmetric braneworld background. The inherent $tr$-symmetry of the background ensures the separability of the perturbation equation, making the problem analytically tractable. The results obtained within this framework are compared with those from the DF approximation, also highlighted in purple colour in the Table (\ref{tab1}).

    \item Importantly, in both the DF approximation and MTE analysis, we have restricted our attention to small values of $Q$. This is physically meaningful, as it ensures that the deformation from the Schwarzschild geometry remains perturbative and the underlying assumptions of the framework continue to hold consistently. Moreover, the constraint on $Q$ obtained from both approaches is found to be compatible, demonstrating the effectiveness of the DF approximation in providing reliable order-of-magnitude bounds on the tidal charge. However, for cases involving higher orbital eccentricities, the difference between mismatches (from the reference waveform) increases (discussed in Sec. (\ref{wave})) for the MTE case, highlighting the importance of the MTE framework for achieving precision-level computations of GW observables.

\end{itemize}

Before ending this section, we comment on the choice of the range of the $Q$ that we make in this paper. Motivated by the LIGO-Virgo-KAGRA (LVK) constraints on braneworld black holes, the allowed range of the tidal charge $Q$ has been tested against GWTC-3 observations \cite{LIGOScientific:2021sio, Kumar:2025njz}. As shown in \cite{Kumar:2025njz}, following the post-Newtonian (PN) analyses of \cite{Glampedakis:2002ya, Ryan:1995xi, Flanagan:2007tv, AbhishekChowdhuri:2023gvu}, the leading-order effect of $Q$ enters at 1PN order. For a binary with mass-ratio $q=0.6$ and a mass $M=140M_\odot$ in the LVK band, the dephasing induced by $Q\in(10^{-6},0.3)$ lies in the range $\delta\Phi\sim10^{-9}$-$10^{-4}$, consistent with the reported 1PN deviation $\delta\Phi_2\sim0.05$ \cite{LIGOScientific:2021sio} (otherwise the effect of this $Q$ would have been already detected by LVK observations). For $Q<10^{-6}$, the dephasing is even smaller, ensuring compatibility with present GW observations. While the analysis of \cite{Kumar:2025njz} assumes $\delta E_{\mu\nu}=0$, the consistency check with the LVK is adequate at the PN level, and the same holds in the context of the present article. Therefore, the dephasing for different $Q$ values (as our focus is to get an estimate for the lowest possible value of $Q$ that can be detected via LISA), we mainly focus on the range $Q\in(10^{-7},10^{-5})$, considered throughout the analysis, remains consistent with those inferred from LVK GWTC-3 data \cite{LIGOScientific:2021sio} for mass-ratios relevant to stellar-mass black holes.

With this, let us look at how the draft is organized. In Section~(\ref{BH}), we describe the spherically symmetric black hole carrying tidal charge and provide expressions for orbital dynamics. Section~(\ref{perturbation}) mentions perturbation equations derived from the MTE framework along with the methods used for GW fluxes, orbital evolution, waveform, and mismatch. In Section~(\ref{wave}), we provide details of the results of dephasing and mismatch showing the importance of the MTE framework for precision computation of the GW observables, and a comparison with the results derived from the DF-Teukolsky formalism carrying modified $\Delta (r)$. Finally, in Section~(\ref{dscn}), we conclude our discussion by sharing insights on key aspects and future prospects.
\par
\textit{Notation and Convention: } We set the fundamental constants $G$ and $c$ to unity and adopt sign convention $(-1,1,1,1)$. The \textit{superscript} and \textit{subscript} indices of Greek and Latin letters ($\mu, \nu, \alpha, \beta, ..., a, b, c, d, ... $) refer to four-dimensional spacetime components.

\section{Braneworld Spacetime and Geodesic Motion}\label{BH}
To derive black hole solutions on the brane, one typically starts from the five-dimensional Einstein field equations (EFEs), describing the geometry and {higher-dimensional bulk spacetime \cite{Chamblin:2000ra, Dadhich:2000am, Shiromizu:1999wj}. By projecting these equations onto the four-dimensional brane, which embeds our observable universe in the bulk, one effectively translates the bulk dynamics into four-dimensional terms. Such a projection brings the notion of extra dimensions through the additional terms in the brane equations. Consequently, the effects of the higher-dimensional geometry manifest on the brane without the need to solve the full bulk spacetime directly. This establishes a framework for understanding how bulk gravitational phenomena can influence the physics we observe in four dimensions, particularly in the context of localized gravitational objects, such as black holes. The induced metric on the brane looks $g_{\mu\nu}=\tilde{g}_{\mu\nu}-n_{\mu} n_{\nu}$, where $\tilde{g}_{\mu\nu}$ is the bulk metric and $n_{\mu}$ is the spacelike unit normal to the brane. The field equations induced on the brane follow from the Gauss-Codazzi relations and the $Z_{2}$-symmetric matching conditions, and can be written as a modification of the standard Einstein field equations \cite{Shiromizu:1999wj, Dadhich:2000am}
\begin{align}
G_{\mu\nu} = -\Lambda g_{\mu\nu} + 8\pi \tau_{\mu\nu} + \tilde{\kappa}^{4} \Pi_{\mu\nu} - E_{\mu\nu},
\end{align}
where $G_{\mu\nu}$ is the Einstein tensor and $E_{\mu\nu}$ represents the projected electric part of the bulk Weyl tensor ($E_{\mu\nu}=\tilde{C}_{\mu\alpha\nu\beta}n^{\alpha}n^{\beta}$; with $\tilde{C}_{\mu\alpha\nu\beta}$ being the bulk Weyl tensor), which comes from the fact that the Weyl tensor can be decomposed into two parts: electric and magnetic, with magnetic part being zero on the brane \cite{Shiromizu:1999wj, Dadhich:2000am}. $E_{\mu\nu}$ carries nonlocal gravitational information from the higher-dimensional bulk, further reflecting the influence of the gravitational field extending beyond the brane \cite{Shiromizu:1999wj, Maartens:2001jx}. $\Pi_{\mu\nu}$ is a quadratic combination of the energy-momentum tensor, including terms such as $(\tau^\alpha_\mu \tau_{\alpha\nu})$ and $(\tau \tau_{\mu\nu})$. $\tilde{\kappa}$ relates to the 5-dimensional Planck mass ($\tilde{M}_{P}$) via $\tilde{\kappa}^{2}=8\pi/\tilde{M}^{3}_{P}$. These contributions, together with the traceless projected Weyl tensor $E_{\mu\nu}$, represent nontrivial corrections to the standard Einstein field equations in GR, encoding local and nonlocal bulk effects on the brane. We consider vacuum solutions, i.e. $\Pi_{\mu\nu}=0=\tau_{\mu\nu}$, together with cosmological constant $\Lambda =0$. The effective filed equations on vacuum reduces to $G_{\mu\nu}+E_{\mu\nu}=0$; $R_{\mu}{}^{\mu} = 0 = E_{\mu}{}^{\mu}$ \cite{Dadhich:2000am}. Additionally, Weyl symmetries require that $E_{\mu\nu}$ be symmetric and traceless. In the vacuum case on the brane, $E_{\mu\nu}$ obeys the divergence constraint: $\nabla^{\mu}E_{\mu\nu}=0$ \cite{Dadhich:2000am, Shiromizu:1999wj}; where $\nabla$ is the brane covariant derivative. Thus, for a vacuum region outside a mass localized on the brane, the solution must satisfy these equations. With this, one can arrive at the  static spherically symmetric metric of the following form \cite{Dadhich:2000am, Rahman:2022fay}:
\begin{align}\label{metric}
ds^{2} = -\frac{\Delta(r)}{r^{2}}dt^{2}+\frac{r^{2}}{\Delta(r)}dr^{2}+r^{2}(d\theta^{2}+\sin^{2}\theta d\phi^{2}).
\end{align}
This metric is a non-Ricci-flat, Petrov type D, and also termed $tr$-symmetric spacetime with $\Delta(r)=r^{2}-2Mr+QM^{2}$ where $Q$ denotes the tidal charge. It can take both positive and negative values. The parameter $M$ is the mass of the black hole. It is apparent that $Q=0$ reproduces the Schwarzschild geometry as well as $Q\rightarrow Q^{2}$ represents the Reissner-Nordstr$\Ddot{\textup{o}}$m (RN) spacetime \cite{Dafermos:2003wr, PhysRevLett.67.789, PhysRevLett.120.031103}. The location of the horizon is given by: $\Delta(r)=0 \Rightarrow r_{\pm} = M(1\pm \sqrt{1-Q})$.

We analyze the evolution of an inspiralling object in an EMRI system that generates GWs, valid for a small mass-ratio ($q=m_{\textup{SO}}/M$). In other words, the inspiralling object generates a small perturbation on the background, and the resulting field equations on the brane can be linearized. The perturbed field equation in vacuum is written as
\begin{align}
    \delta G_{\mu\nu} + \delta E_{\mu\nu} = 0.
\end{align}
As mentioned earlier that $\Pi_{\mu\nu}$ is quadratic in mass-ratio ($q^{2}$), its contribution can be neglected anyway at the linear order in $q$. Additionally, in the low-energy limit, the perturbation of the projected Weyl tensor is negligible ($\delta E_{\mu\nu}=0$), as stated earlier, which is justified when the brane energy density is much smaller than the brane tension. We have analyzed EMRIs under such a DF approximation in \cite{Rahman:2022fay, Kumar:2025njz, Zi:2024dpi} in the limit of small $Q$ (tidal charge), which captures the non-GR physics. In the present study, we aim to relax this assumption and examine the effect of non-Ricci-flatness appearing from such a non-vanishing contribution of $\delta E_{\mu\nu}\,.$ At the level of the perturbation equation, in Section (\ref{app:MTE}), we notice that the notion of such an effect enters in the modified Teukolsky potential by the Ricci scalar $\Phi_{11}$. Briefly, we consider the effective description of the metric localized on the brane from the 4D point of view and implement the MTE framework to incorporate the nonzero $\Phi_{11}$, indicating the non Ricci-flat nature of the brane black hole.

The spacetime in Eq. (\ref{metric}) admits two symmetries, leading to conserved quantities: particle energy $E$ and angular momentum $L_{z}$. These constants can be used to derive the general form of the particle's timelike four-velocity, yielding the following expression:
\begin{equation}
\begin{aligned}\label{geodesic}
m_{\textup{SO}}\frac{dt}{d\tau} =& \frac{E r^2}{\Delta(r)}; \hspace{7mm}
m_{\textup{SO}}\frac{d\phi}{d\tau} = \frac{L_{z}}{r^{2}}, \\
m_{\textup{SO}}^{2}\Big(\frac{dr}{d\tau}\Big)^{2} =& E^{2}-\frac{\Delta(r)}{r^{2}}\Big(m_{\textup{SO}}^{2}+\frac{L_z^2}{r^{2}}\Big).
\end{aligned}
\end{equation}
Here, $m_{\textup{SO}}$ denotes the mass of the inspiraling object (secondary object). For simplicity, we adopt dimensionless units defined by $\hat{r}=r/M$, $\hat{E}=E/m_{\textup{SO}}$ and $\hat{L}_{z}=L_{z}/(m_{\textup{SO}} M)$ \cite{PhysRevD.102.024041}. For writing convenience, we omit the hats in later expressions of the article with the understanding that all quantities are expressed in these rescaled units. However, physical units can be restored at any stage if needed. As we are taking equatorial motion, we set $\theta=\pi/2$ \cite{Glampedakis:2002cb, Glampedakis:2002ya, PhysRevD.61.084004}. Furthermore, we know that eccentric orbits are characterized by two turning points: the periastron ($r_{p}=\frac{p}{1+e}$) and the apastron ($r_a=\frac{p}{1-e}$), with motion confined to the range $r_p<r<r_a$, where the effective potential $V_{\textup{eff}}(r)<0$ is negative \cite{PhysRevD.103.104045}. To avoid singular behavior in solving various differential equations at these turning points, it is useful to reparametrize the orbits in a more general form to determine the orbital motion by eliminating the dependence of proper time ($\tau$) \cite{PhysRevD.50.3816}. We adopt the following orbital parametrization:
\begin{align}\label{parametrize}
r = \frac{p}{1+e\cos\chi},
\end{align}
where $\chi\in (0, 2\pi)$. With the use of Eq. (\ref{geodesic}) and Eq. (\ref{parametrize}) and considering the turning points in eccentric dynamics, one can obtain the orbital energy and angular momentum, as well as the last stable orbit (LSO) or separatrix ($p_{\textup{sp}}$) location. The full analytical expressions for obtaining these quantities are highly non-trivial as the intermediate structure of the expressions is very cumbersome and also such expressions are not required for the objectives of our analysis, so we focus only on the numerical evaluation of these quantities,  which is sufficient for computing the relevant physical quantities. Although the scheme for linear correction (say in the separatrix) in any non-GR parameter ($Q$ in our case) is well established in the literature \cite{Cutler:1994ys, Glampedakis:2002ya, Kumar:2024utz, AbhishekChowdhuri:2023gvu}, allowing for an equivalent separatrix expression, we do not linearize these quantities in $Q$ throughout our analysis. Using numerical values thus provides a more practical and accurate approach.

Further, the eccentric orbits exhibit two fundamental frequencies: azimuthal ($\Omega_{\phi}$) and radial ($\Omega_{r}$).
This allows us to track the orbital phase while accounting for the influence of the tidal charge parameter. Since obtaining closed-form expressions for these quantities is analytically challenging, we rely on their numerical values in the results presented later in the paper. We obtain the dephasing in the following way \cite{PhysRevD.50.3816}
\begin{align}
\delta\Phi_{\phi,r} = 2\int_{0}^{t}(\Omega^{Q\neq 0}_{\phi,r}-\Omega^{Q =0}_{\phi,r})dt\,.
\end{align}
A dephasing of about 1 rad in the radial or azimuthal phase is typically used as the threshold for LISA to distinguish between two EMRI signals \cite{PhysRevD.101.044004}. We will analyze and compare the results with the impact of distinct values of ($Q, e$), both in the case of the DF Teukolsky and MTE.

\section{Modified Teukolsky Perturbation and Method} \label{perturbation}
In this section, we explore the decoupled, gauge-invariant equation governing perturbations of a spherically symmetric, asymptotically flat spacetime, extending beyond GR. Building on this foundation, we can obtain a simple form of the MTE for spacetimes exhibiting a $tr$-symmetry and non-Ricci-flatness \cite{Li:2022pcy, Cano:2023tmv, Guo:2024bqe, Guo:2023wtx, Arbey:2021jif}, where the geometry can be treated as a linear perturbation of a Petrov type D spacetime in GR. We further provide the details for obtaining the decoupled MTE.

\subsection{The MTE}\label{app:MTE}\vspace{-0.2cm}
Here, we provide the NP tetrad, spin-coefficients, and relevant expressions for the MTE for the spacetime under consideration \cite{Newman:1961qr}. The readers are suggested to take a look at the full construction in \cite{Li:2022pcy, Guo:2023wtx, Guo:2024bqe, Arbey:2021jif}
\begin{equation}
\begin{aligned}\label{NP1}
l^{a} =& \frac{1}{\Delta(r)}\Big[r^{2}, \Delta(r), 0, 0 \Big] \hspace{2mm} ; \hspace{2mm}
n^{a} = \frac{1}{2r^{2}}\Big[r^{2}, -\Delta(r), 0, 0 \Big], \\
m^{a} =& \frac{1}{r\sqrt{2}}\Big[0, 0, 1, i\csc\theta \Big] \hspace{2mm} ; \hspace{2mm}
\bar{m}^{a} = \frac{1}{r\sqrt{2}}\Big[0, 0, 1, -i\csc\theta \Big],
\end{aligned}
\end{equation}

\begin{equation}
\begin{aligned}\label{NP2}
\rho =& \frac{1}{r} ; \hspace{1mm} \mu = \frac{\Delta(r)}{2r^{3}} ; \hspace{1mm} \gamma = \frac{2\Delta(r)-r\Delta'(r)}{4r^{3}}, \\
 -\alpha =& \beta = \frac{\cot\theta}{2r\sqrt{2}};\hspace{1mm} \Psi_{2} = \frac{6r\Delta'(r)-12\Delta(r)}{12r^{4}}, \\
 \Phi_{11} =& \frac{4\Delta(r)+r(-4\Delta'(r)+r(2+\Delta''(r)))}{8r^{4}},
\end{aligned}
\end{equation}
where $\Phi_{11}=\frac{1}{4}R_{ab}(l^{a}n^{b}+m^{a}\bar{m}^{b})$, characterizing the non-Ricci-flat nature of the braneworld geometry and the Weyl scalar $\Psi_{2}=C_{abcd}l^{a}m^{b}\bar{m}^{c}n^{d}$. ($\rho, \mu, \gamma, \alpha, \beta$) are spin coefficients, defined as: $\rho=-m^{a}\bar{m}^{b}\nabla_{b}l_{a}$, $\mu=m^{a}\bar{m}^{b}\nabla_{b}n_{a}$, $\gamma=-\frac{1}{2}(n^{a}n^{b}\nabla_{b}l_{a}-\bar{m}^{a}n^{b}\nabla_{b}m_{a})$, $\alpha=-\frac{1}{2}(n^{a}\bar{m}^{b}\nabla_{b}l_{a}-\bar{m}^{a}\bar{m}^{b}\nabla_{b}m_{a})$ and $\beta=-\frac{1}{2}(n^{a}m^{b}\nabla_{b}l_{a}-\bar{m}^{a}m^{b}\nabla_{b}m_{a})$; with $\nabla_{b}$ being the covariant derivative with respect to the background. Let us now briefly look at the gravitational perturbation equation for the MTE case. We start with the NP formalism \cite{Newman:1961qr}:
\begin{widetext}
\begin{equation}\label{eqn1}
\begin{aligned}
(\delta + 4\beta - \tau)\,\Psi_4
- (\Delta + 4\mu + 2\gamma)\,\Psi_3
+ 3\nu\,\Psi_2
=
(\bar\delta - \bar\tau + 2\bar\beta + 2\alpha)\,\Phi_{22} &
- (\Delta + 2\gamma + 2\bar\mu)\,\Phi_{21}
- 2\lambda\,\Phi_{12}
+ 2\nu\,\Phi_{11}
+ \bar\nu\,\Phi_{20} \\
(D + 4\epsilon - \rho)\,\Psi_{4}
- (\bar{\delta} + 4\pi + 2\alpha)\,\Psi_{3}
+ 3\lambda\,\Psi_{2}
=
(\bar{\delta} - 2\bar{\tau} + 2\alpha)\,\Phi_{21} &
- (\Delta + 2\gamma - 2\bar{\gamma} + \bar{\mu})\,\Phi_{20}
+ \bar{\sigma}\,\Phi_{22}
- 2\lambda\,\Phi_{11}
+ 2\nu\,\Phi_{10} \\
(\Delta + \mu + \bar{\mu} + 3\gamma - \bar{\gamma})\,\lambda
- (\bar{\delta} + \pi - \bar{\tau} + \bar{\beta} + 3\alpha)\,\nu
+ \Psi_{4} =& 0\,.
\end{aligned}
\end{equation}
\end{widetext}
We decompose all NP scalar quantities into a background contribution as $\Psi_{j}=\Psi^{(0)}_{j}+\chi\Psi^{(1)}_{j}$ with (0) being the background and (1) representing the first-order perturbation. We keep the parameter $\chi$ as a linear perturbation of NP scalars, and set $\chi=1$ for linearized perturbation. Substituting this decomposition into Eq. (\ref{eqn1}), one finds that the background pieces (0) are already satisfied by construction. The equations that remain at first order therefore govern the perturbations, i.e., the terms carrying the superscript (1). With this, the first-order perturbation, in the NP formalism, can be written as
\begin{widetext}
\begin{equation}\label{eq1}
\begin{aligned}
& (\delta+4\beta-\tau)^{(0)}\Psi_{4}^{(1)}-(\tilde{\Delta}+4\mu+2\gamma)^{(0)}\Psi_{3}^{(1)}+3\nu^{(1)}\Psi^{(0)}_{2}
=(\bar{\delta}-\bar{\tau}+2\bar{\beta}+2\alpha)^{(0)}\Phi_{22}^{(1)}
-(\tilde{\Delta}+2\gamma+2\bar{\mu})^{(0)}\Phi_{21}^{(1)}+2\nu^{(1)}\Phi^{(0)}_{11}
\\
& (D+4\epsilon-\rho)^{(0)}\Psi_{4}^{(1)}-(\bar{\delta}+4\pi+2\alpha)^{(0)}\Psi_{3}^{(1)}+3\lambda^{(1)}\Psi^{(0)}_{2} =(\bar{\delta}-2\bar{\tau}+2\alpha)^{(0)}\Phi_{21}^{(1)}-(\tilde{\Delta}+2\gamma-2\bar{\gamma}+\bar{\mu})^{(0)}\Phi_{20}^{(1)}
-2\lambda^{(1)}\Phi^{(0)}_{11}
\\
& (\tilde{\Delta}+\mu+\bar{\mu}+3\gamma-\bar{\gamma})^{(0)}\lambda^{(1)}-(\bar{\delta}+\pi-\bar{\tau}+\bar{\beta}+3\alpha)^{(0)}\nu^{(1)}
+\Psi_{4}^{(1)}=0\,.
\end{aligned}
\end{equation}
\end{widetext}
Further, we observe that the following commutation relation between the differential operators holds true not only in vacuum, but also in the background under consideration:
\begin{widetext}
\begin{equation}\label{eq2}
\begin{aligned}
&(\tilde{\Delta}+(p+1)\gamma-\bar{\gamma}-q\mu+\bar{\mu})^{(0)}(\bar{\delta}+p\alpha-q\pi)^{(0)} -(\bar{\delta}+(p+1)\alpha+\bar{\beta}-\bar{\tau}-q\pi)^{(0)}(\tilde{\Delta}+p\gamma-q\mu)^{(0)} \\
&=\nu^{(0)} D^{(0)}-\lambda^{(0)}\delta^{(0)}-p((\beta+\tau)\lambda-(\rho+\epsilon)\nu+\Psi_{3})^{(0)} +q[-D\nu+\delta\lambda+(\bar{\pi}+\tau+3\beta-\bar{\alpha})\lambda -(3\epsilon+\bar{\epsilon}+\rho-\bar{\rho})\nu+2\Psi_{3}]^{(0)} \\
&=0\,,
\end{aligned}
\end{equation}
\end{widetext}
where $(p, q)$ are two constants. At this point imposing the conditions $\lambda^{(0)} = \nu^{(0)} =\Psi^{(0)}_{3} = 0$, and applying the differential operators $(\tilde{\Delta}+3\gamma-\bar{\gamma}+4\mu+\bar{\mu})^{(0)}$ and $(\bar{\delta}+3\alpha+\bar{\beta}-\bar{\tau}+4\pi)^{(0)}$ to the second equation of Eq. (\ref{eq1}) and first equation of Eq. (\ref{eq1}), respectively. Further, subtracting first from the second in Eq. (\ref{eq1}), we find that the terms involving $\Psi^{(1)}_{3}$ drop out as a consequence of the commutator identity in Eq. (\ref{eq2}), evaluated with $(p, q) = (2, -4)$. What we obtain is the following
\begin{widetext}
\begin{equation}\label{eq4}
\begin{aligned}
&\left[(\tilde{\Delta}+3\gamma-\bar{\gamma}+4\mu+\bar{\mu})(D+4\epsilon-\rho)
-(\bar{\delta}+3\alpha+\bar{\beta}-\bar{\tau}+4\pi)(\delta+4\beta-\tau)\right]^{(0)}\Psi_{4}^{(1)} \\
& +(3\Psi_{2}+2\Phi_{11})^{(0)}(\Delta+3\gamma-\bar{\gamma}+4\mu+\bar{\mu})^{(0)}\lambda^{(1)}
 -(3\Psi_{2}-2\Phi_{11})^{(0)}(\bar{\delta}+3\alpha+\bar{\beta}-\bar{\tau}+4\pi)^{(0)}\nu^{(1)} \\
&=T_{4} -\lambda^{(1)}\tilde{\Delta}^{(0)}(3\Psi_{2}+2\Phi_{11})^{(0)}+\nu^{(1)}\bar{\delta}^{(0)}(3\Psi_{2}-2\Phi_{11})^{(0)},
\end{aligned}
\end{equation}
\end{widetext}
where $T_{4}$ is defined by
\begin{widetext}
\begin{equation}
\begin{aligned}
T_{4}=&(\tilde{\Delta}+3\gamma-\bar{\gamma}+4\mu+\bar{\mu})^{(0)}
\left[(\bar{\delta}-2\bar{\tau}+2\alpha)^{(0)}\Phi_{21}^{(1)}
-(\tilde{\Delta}+2\gamma-2\bar{\gamma}+\bar{\mu})^{(0)}\Phi_{20}^{(1)}\right]-(\bar{\delta}+3\alpha+\bar{\beta}-\bar{\tau}+4\pi)^{(0)} \\
& \left[(\bar{\delta}-\bar{\tau}+2\bar{\beta}+2\alpha)^{(0)}\Phi_{22}^{(1)}
-(\tilde{\Delta}+2\gamma+2\bar{\mu})^{(0)}\Phi_{21}^{(1)}\right].
\end{aligned}
\end{equation}
\end{widetext}
Now, using the following relations on the right-hand side of Eq. (\ref{eq4}): \\
\\
\begin{widetext}
\begin{equation}
\begin{aligned}
\tilde{\Delta}^{(0)}(3\Psi_{2}-2\Phi_{11})^{(0)}=&-3\mu^{(0)}(3\Psi_{2}+2\Phi_{11})^{(0)}+4(\mu+\bar{\mu})^{(0)}\Phi^{(0)}_{11}, \\
\bar{\delta}^{(0)}(3\Psi_{2}+2\Phi_{11})^{(0)}=&-3\pi^{(0)}(3\Psi_{2}-2\Phi_{11})^{(0)}-4(\pi-\bar{\tau})^{(0)}\Phi^{(0)}_{11}.
\end{aligned}
\end{equation}
\end{widetext}
As a result, we get, \vspace{1cm}
\begin{widetext}
\begin{equation}\label{eq6}
\begin{aligned}
&[(\tilde{\Delta}+3\gamma-\bar{\gamma}+4\mu+\bar{\mu})(D+4\epsilon-\rho)
-(\bar{\delta}+3\alpha+\bar{\beta}-\bar{\tau}+4\pi)(\delta+4\beta-\tau)-3\Psi_{2}]^{(0)}\Psi_{4}^{(1)}+2\Phi^{(0)}_{11}(\tilde{\Delta}+3\gamma-\bar{\gamma}+\mu+\bar{\mu})^{(0)}\lambda^{(1)} \\
&+2\Phi^{(0)}_{11}(\bar{\delta}+3\alpha+\bar{\beta}-\bar{\tau}+\pi)^{(0)}\nu^{(1)} =T_{4} -4\lambda^{(1)}[(\tilde{\Delta}+\mu+\bar{\mu})\Phi_{11}]^{(0)}-4\nu^{(1)}[(\bar{\delta}+\pi-\bar{\tau})\Phi_{11}]^{(0)}.
\end{aligned}
\end{equation}
\end{widetext}
In the vacuum case, Eq. (\ref{eq6}) leads to a decoupled expression for $\Psi^{(1)}_{4}$ as $\Phi_{11}=0$. By decoupled, we refer to an equation that is independent of other perturbed Weyl scalars and spin coefficients. Moreover, the current background still involves $(\nu^{(1)}, \lambda^{(1)})$, which can be removed through appropriate gauge conditions. A possible gauge choice is to set $\Psi^{(1)}_{3}=0$ which enables both ($\lambda^{(1)}, \nu^{(1)}$) to be expressed in terms of $\Psi^{(1)}_{4}$. However, the wave equation becomes complicated; in this case, one might have to look for other gauge choices. Alternatively, we can eliminate either $\nu^{(1)}$ or $\lambda^{(1)}$. Eq. (\ref{eq6}) gets simplified once the $\nu^{(1)}$ is eliminated, further using the gauge condition $(\tilde{\Delta}+3\gamma-\bar{\gamma}+2\mu+2\bar{\mu})^{(0)}(\Phi^{(0)}_{11}\lambda^{(1)})+\nu^{(1)}[(\bar{\delta}+\pi-\bar{\tau})\Phi_{11}]^{(0)}=0$, the decoupled equation for $\Psi^{(1)}_{4}$ is given by \cite{Li:2022pcy, Guo:2023wtx, Guo:2024bqe, Arbey:2021jif}
\begin{widetext}
\begin{equation}\label{eq7}
\begin{aligned}
&\left[(\tilde{\Delta}+3\gamma-\bar{\gamma}+4\mu+\bar{\mu})(D+4\epsilon-\rho)-(\bar{\delta}+3\alpha+\bar{\beta}-\bar{\tau}+4\pi)(\delta+4\beta-\tau)-3\Psi_{2}+2\Phi_{11}\right]^{(0)}\Psi_{4}^{(1)}=T_{4}\,.
\end{aligned}
\end{equation}
\end{widetext}
This is the perturbation equation for the spherical $tr$-symmetric static background.
where ($\gamma, \mu, \delta, \beta, \tau, \pi, \alpha$) are NP quantities; ($\Tilde{\Delta}, D, \delta, \bar{\delta}$) are derivative operators. $T_{4}$ is the source term, which is, in the Teukolsky framework, written with the use of the NP tetrad basis, modeling it as a point particle: 
\begin{align}
    T^{\mu\nu} = m_{\rm SO} \int\frac{d\tau}{\sqrt{-g}}\frac{dz^{\mu}}{d\tau}\frac{dz^{\nu}}{d\tau}\delta^{(4)}(x-z(\tau))\,.
\end{align}

As noted, the presence of $\Phi_{11}$ reflects the non-Ricci-flat nature of the braneworld spacetime, indicating that $\delta E_{\mu\nu} \neq 0$. In our analysis, we consider this contribution to be arising solely from the non-Ricci-flatness; however, a full description of $E_{\mu\nu}$, possibly may lead to additional physics, requires a systematic inclusion of those additional correction, indicating requirement of a  more general treatment, as described in \cite{Li:2022pcy}, which is beyond the current scope of the present article and remains the interest of future investigations. Our work provides a first step toward exploring non-Ricci-flat, Petrov type D spacetimes--particularly, spherically symmetric braneworld black holes--where the $tr$-symmetry of the metric allows a compact formulation of the master perturbation equations, MTE. The analysis also compares the existing results based on the assumption $\delta E_{\mu\nu}=0$ adopted in previous studies \cite{Kumar:2025njz, Zi:2024dpi, Rahman:2022fay} based on the DF approximation. 

Using the expressions mentioned in Eqs. (\ref{NP1}), (\ref{NP2}) and (\ref{eq7}), we can write down the separable radial and angular equations in the following fashion:
\begin{equation}
\begin{aligned}\label{radequ}
\Delta^{-s}\frac{d}{dr}\Big(\Delta^{s+1}\frac{dR_{s}}{dr} \Big) + \Big[\frac{K(K-is\Delta')}{\Delta}+4is\omega r-\lambda \\
-\varepsilon V_{s}(r) \Big] R_{s} = \mathcal{T}_{s\ell m\omega}\,, \\
\Big[\frac{1}{\sin\theta}\frac{d}{d\theta}\Big(\sin\theta\frac{d}{d\theta}\Big)-\Big(\frac{m+s\cos\theta}{\sin\theta}\Big)^{2}+s+\lambda_{s}\Big] {}_{s}S_{\ell m} = 0\,,
\end{aligned}
\end{equation}
where $K=r^{2}\omega$ and $s\in (0, 1, \pm 2)$ with $s=-2$ for gravitational perturbation for outgoing waves in our context. $\lambda_{s} = (\ell-s)(\ell+s+1)$ is the eigenvalue of the spin-weighted spherical harmonics (${}_{s}S_{\ell m}$). $\mathcal{T}_{slm\omega}$ is the source term of the inspiraling object, described in the appendix (\ref{source}). The angular equation takes the same form as it appears for the Schwarzschild case. $R_{s}(r)$ is the radial function. Importantly, the radial part of the equation has contributing terms beyond GR, which are the coefficients of a bookkeeping parameter $\varepsilon$ with $V_{s}(r) =\frac{2\Delta'}{r}-\frac{2\Delta}{r^{2}}-\frac{s+4}{2}\Delta''+(s+2)$. This correction, derived from the MTE, represents the contribution from the non-Ricci-flatness nature of the spacetime, indicating a partial contribution from the underlying theory that comes from the bulk Weyl tensor. Note that $\varepsilon$ can take either 0 or 1; if $\varepsilon=0$, it implies the results from the DF-Teukolsky equation with modified $\Delta (r)$; if $\varepsilon=1$, it indicates the results with the contribution of MTE corrections.


Now with the Eq. (\ref{radequ}), the homogeneous solution of the radial equation is given by,
\begin{equation}
\begin{aligned}\label{rty1}
  R^{in}_{\ell m\omega}(r)\sim & \begin{cases}
    B^{tran}_{\ell m\omega}\Delta^{2}e^{-ik r_{*}}; & \text{$r\longrightarrow  r_{+}$},\\
    B^{out}_{\ell m\omega}r^{3}e^{i\omega r_{*}}+B^{in}_{\ell m\omega}r^{-1}e^{-i\omega r_{*}}; & \text{$r\longrightarrow \infty$},
  \end{cases}\\
  R^{up}_{\ell m\omega}(r)\sim & \begin{cases}
    C^{out}_{\ell m\omega}e^{ik r_{*}}+C^{in}_{\ell m\omega}\Delta^{2}e^{-ikr_{*}}, & \text{$r\longrightarrow r_{+}$},\\
    C^{tran}_{\ell m\omega}r^{3}e^{i\omega r_{*}}; & \text{$r\longrightarrow \infty$},
  \end{cases}
  \end{aligned}
\end{equation}
where $k = \omega$ and $\frac{dr_{*}}{dr} = \frac{r^{2}}{\Delta}$. 
The focus of this study is to provide a comparative estimate of the constraints on the tidal charge parameter obtained from both the DF Teukolsky equation and the MTE.

Since the nature of perturbation equations is very similar to the DF-Teukolsky case, indicating the additional changes in the Teukolsky potential, we implement the Sasaki-Nakamura (SN) method \cite{10.1143/PTP.67.1788, SASAKI198185, SASAKI198268, KOJIMA1983335}, which recasts the perturbation equation into a new form, known as the SN equation, featuring a short-range potential, which makes it more suitable for efficient numerical treatment. Following \cite{PhysRevD.102.024041, 10.1143/PTP.67.1788, Kumar:2025njz}, the radial equation can be recast into a form of the SN equation. To extract the asymptotic coefficients of this transformed equation, we apply the Frobenius method, expanding the solution as a power series. We compute mismatch and consequently compare the constraint on the tidal charge between the DF-Teukolsky ($\varepsilon = 0$) and MTE ($\varepsilon = 1$) cases.

\subsection{Flux and Evolution}
Following \cite{PhysRevD.102.024041, Sasaki:2003xr, Hughes:1999bq, Datta:2024vll, Barsanti:2022ana, Kumar:2025njz}, energy and angular momentum fluxes computed at infinity and horizon for an EMRI system can be written as

\begin{equation}\label{fluxes}
\begin{aligned}
\Big(\frac{dE}{dt}\Big)^{\infty}_{\rm GW}  =& \sum_{\ell mn} \frac{|Z^{H}_{\ell mn}|^2}{4\pi \omega_{\ell mn}^2}\,, \\
\Big(\frac{dE}{dt}\Big)^{H}_{\rm GW} =& \sum_{\ell mn}\alpha_{\ell mn} \frac{|Z^{\infty}_{lmn}|^2}{4\pi \omega_{\ell mn}^2}\,, \\
\Big(\frac{dL_z}{dt}\Big)^{\infty}_{\rm GW}  =& \sum_{\ell mn} \frac{m |Z^{H}_{\ell mn}|^2}{4\pi \omega_{\ell mn}^3}\,,\\
\Big(\frac{dL_z}{dt}\Big)^{H}_{\rm GW} =& \sum_{\ell mn} \frac{\alpha_{\ell mn} m|Z^{\infty}_{lmn}|^2}{4\pi \omega_{\ell mn}^3}\,,
\end{aligned}
\end{equation}
where
\begin{align}\label{ampl}
Z^{H}_{\ell m\omega} =& \frac{1}{2i\omega B^{in}_{\ell m\omega}} \int_{r_{+}}^{\infty}\frac{R^{in}_{\ell m\omega}}{\Delta^{2}}\mathcal{T}_{\ell m\omega} dr\,, \\ Z^{\infty}_{\ell m\omega} =& \frac{B^{tran}_{\ell m\omega}}{2i\omega B^{in}_{\ell m\omega}C^{tran}_{\ell m\omega}}\int_{r_{+}}^{\infty}\frac{R^{up}_{\ell m\omega}}{\Delta^{2}}\mathcal{T}_{\ell m\omega} dr\,.
\end{align}
The fluxes at both infinity and the horizon are determined by the amplitudes $(Z^{\infty}_{\ell m \omega}, Z^{H}_{\ell m \omega})$. For equatorial orbits, these amplitudes can be expressed as: $
Z^{\infty,H}_{\ell m \omega} = \sum_{n=-\infty}^{\infty}
Z^{\infty,H}_{\ell m n}\delta(\omega - \omega_{mn})$, where $\omega_{mn} = m \Omega_{\phi} + n \Omega_{r}$. We notice that both frequencies, azimuthal and radial ($\Omega_{\phi}, \Omega_{r}$) contribute to the flux computation. The total flux is the sum of fluxes at infinity and horizon: $(d\mathcal{C}/dt)_{\rm GW} = (d\mathcal{C}/dt)^{H}_{\rm GW}+(d\mathcal{C}/dt)^{\infty}_{\rm GW}$ where $\mathcal{C}\in (E, L_z)$. While computing the fluxes, we numerically integrate over $\chi$ that appears from the parametrization (\ref{parametrize}). This implies evaluating fluxes at different orbital radii. One can take different values of tidal charge both in the case of DF-Teukolsky and MTE formalisms, leading to dynamics (fluxes) that are not identical. Note that the results obtained from the DF-Teukolsky equation ($\varepsilon =0$ with modified $\Delta(r)$) are the same as presented in \cite{Kumar:2025njz} if set $a=0$.

We work within the adiabatic approximation, where the inspiral evolves much more slowly than the orbital timescale. This allows us to treat the trajectory as geodesic, with energy and angular momentum losses directly tied to the corresponding fluxes: $(d\mathcal{C}/dt)^{\textup{(orbit)}} = -(d\mathcal{C}/dt)_{\rm GW}, \hspace{0.2cm} \mathcal{C} \in (E, L_z).$ We further use the computed fluxes to determine the adiabatic evolution of the secondary object of the EMRI \cite{Hughes:2021exa}:
\begin{equation}
\begin{aligned}\label{evol1}
\frac{dp}{dt} &= \frac{(1-e^2)}{2}\frac{dr_a}{dt} +\frac{(1+e^2)}{2}\frac{dr_p}{dt}\,,\\
\frac{de}{dt} &= \frac{(1-e^2)}{2p} \Bigg[(1-e) \frac{dr_a}{dt} +(1+e)\frac{dr_p}{dt} \Bigg]\,,
\end{aligned}
\end{equation}
where the expressions for $\frac{dr_{a,p}}{dt}$ can be written as: $\frac{dr_{a,p}}{dt} = \Big(\frac{\partial E}{\partial r_{a,p}}\Big)^{-1}  \left(\frac{dE}{dt}\right)^{\textup{(orbit)}} + \Big(\frac{\partial L_z}{\partial r_{a,p}}\Big)^{-1} \left(\frac{dL_z}{dt}\right)^{\textup{(orbit)}}$. To study the effect of the tidal charge on the orbital evolution, we use the flux-balance law as mentioned above. We construct a two-dimensional grid for fixed values of $Q$ and compute the energy and angular momentum fluxes. The grid includes 40 points in $e\in (0, 0.6)$ and 50 points in $p$, defined as: $p_{i} = p_{\textup{min}}+4(e^{idu}-1)$ with $p_{\textup{min}}=p_{\textup{sp}}+0.02$ and $du = 0.2$ \cite{Datta:2024vll}. This setup ensures a denser grid near the inner edge, improving interpolation. Since quantities vary rapidly near the separatrix, we offset the grid slightly to maintain the validity of the adiabatic approximation. The small exclusion region $p_{\textup{sp}}\leq p \leq p_{\textup{sp}}+0.02$  introduces negligible error, allowing for stable and accurate numerical results in the regimes where the adiabatic method is applicable. It is to note that we only focus on the dominant mode $(\ell, m, n) = (2, 2, 2)$, as including higher modes greatly increases computational cost, especially on the inspiral grid, as also discussed in detail in \cite{Kumar:2025njz}. This mode captures the key features of gravitational radiation in most binary black hole systems \cite{OBrien:2019hcj}, while an analysis of higher modes is left for future work using more efficient methods.
\subsection{Waveform and Mismatch} Once the evolution of the orbital parameters is obtained, the corresponding EMRI waveform is efficiently generated using the quadrupole formula within the \texttt{FastEMRIWaveforms (FEW)} framework \cite{Katz:2021yft, Barack:2003fp, Chua:2020stf}. For massive black holes, \texttt{FEW} employs the \texttt{Augmented Analytical Kludge (AAK)} model to compute the LISA response in the low-frequency regime \cite{Chua:2017ujo}. Thus, with the quadrupolar approximation, \texttt{AAK}  model generates the waveform, where the two polarization modes are represented as a series expansion over harmonic components in the transverse-traceless gauge written as
\begin{align}\label{amplitude}
h_+ \equiv \sum_n A_n^+ = \sum_n &-\Big[1+(\hat{L}\cdot\hat{n})^2\Big]\Big[a_n \cos2\gamma -b_n\sin2\gamma\Big] \nonumber\\
& +c_n\Big[1-(\hat{L}\cdot\hat{n})^2\Big]\,, \\
h_\times \equiv \sum_n A_n^\times = \sum_n &2(\hat{L}\cdot\hat{n})\Big[b_n\cos2\gamma+a_n\sin2\gamma\Big]\;,\nonumber
\end{align}
where $\gamma = \Phi_{\phi} - \Phi_r$ is the direction of eccentric orbit pericenter. ($\hat{n}, \hat{L}$) is the unit vector along the source direction and orbital angular momentum, respectively, and ($a_{n}, b_{n}, c_{n}$) are eccentricity-dependent functions from \cite{Peters:1963ux}.
\begin{equation}
\begin{aligned}
a_n =~ &-n \mathcal{A} \Big[J_{n-2}(ne)-2eJ_{n-1}(ne)+\frac{2}{n}J_n(ne) \\
& +2J_{n+1}(ne) -J_{n+2}(ne)\Big]\cos(n\Phi_r)\,, \\
b_n =~ &-n \mathcal{A}(1-e^2)^{1/2}\Big[J_{n-2}(ne)-2J_{n}(ne)+J_{n+2}(ne)\Big]\\
& \times\sin(n\Phi_r)\,, \\
c_n =~& 2\mathcal{A}J_n(ne)\cos(n\Phi_r) \hspace{1mm} ;
\hspace{1mm} \mathcal{A} = ~ (2\pi M \omega_\phi )^{2/3}m_{\textup{SO}}/d\,,
\end{aligned}
\end{equation}
where $J_n$ denotes Bessel functions of the first kind, with \cite{Barack:2003fp}
\begin{eqnarray}
\hat{L}\cdot\hat{n} = \cos\theta_S\cos\theta_L + \sin\theta_S \sin\theta_L \cos(\phi_S-\phi_L).
\end{eqnarray}

More details about the angles $(\theta_L, \phi_L)$ and $(\theta_K, \phi_K, \Phi_r, \Phi_\phi)$ can be found in \cite{Barack:2003fp}. Thus, the low-frequency EMRI signal, detected by the space-based detector, can be written using full antenna pattern functions ($F^{+,\times}_{I,II}$) \cite{Cutler:1994ys} in the following way
\begin{equation}
h_{\rm I,II} = \frac{\sqrt{3}}{2} (F^+_{\rm I,II} h^+ + F^\times_{\rm I,II} h^{\times})\,.
\end{equation}
More details can be found in \cite{PhysRevD.69.082005}.
Once the waveform is generated, to investigate further the impact of the tidal charge on the EMRI waveforms, we perform a mismatch analysis between waveforms generated with and without the tidal charge parameter. This approach allows us to quantify how significantly the presence of the tidal charge alters the waveform morphology. Given two waveforms, $h_{a}$ (with tidal charge) and $h_{b}$ (without tidal charge), the mismatch is computed using their normalized inner product. The mismatch can be determined through \cite{Cutler:1994ys}
\begin{figure*}[htb!]
\centering
\includegraphics[width=3.2in, height=2.2in]{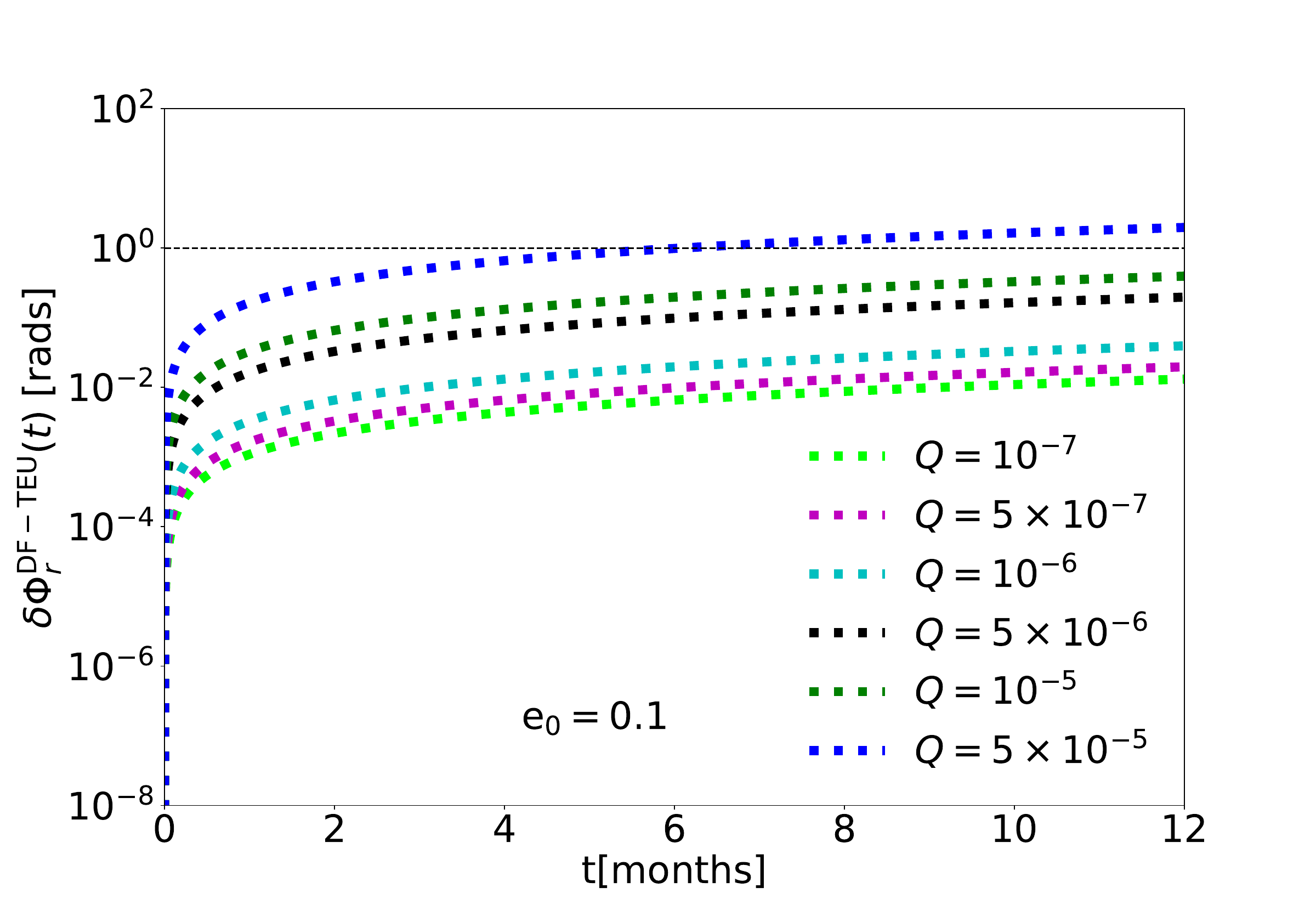}
\includegraphics[width=3.2in, height=2.2in]{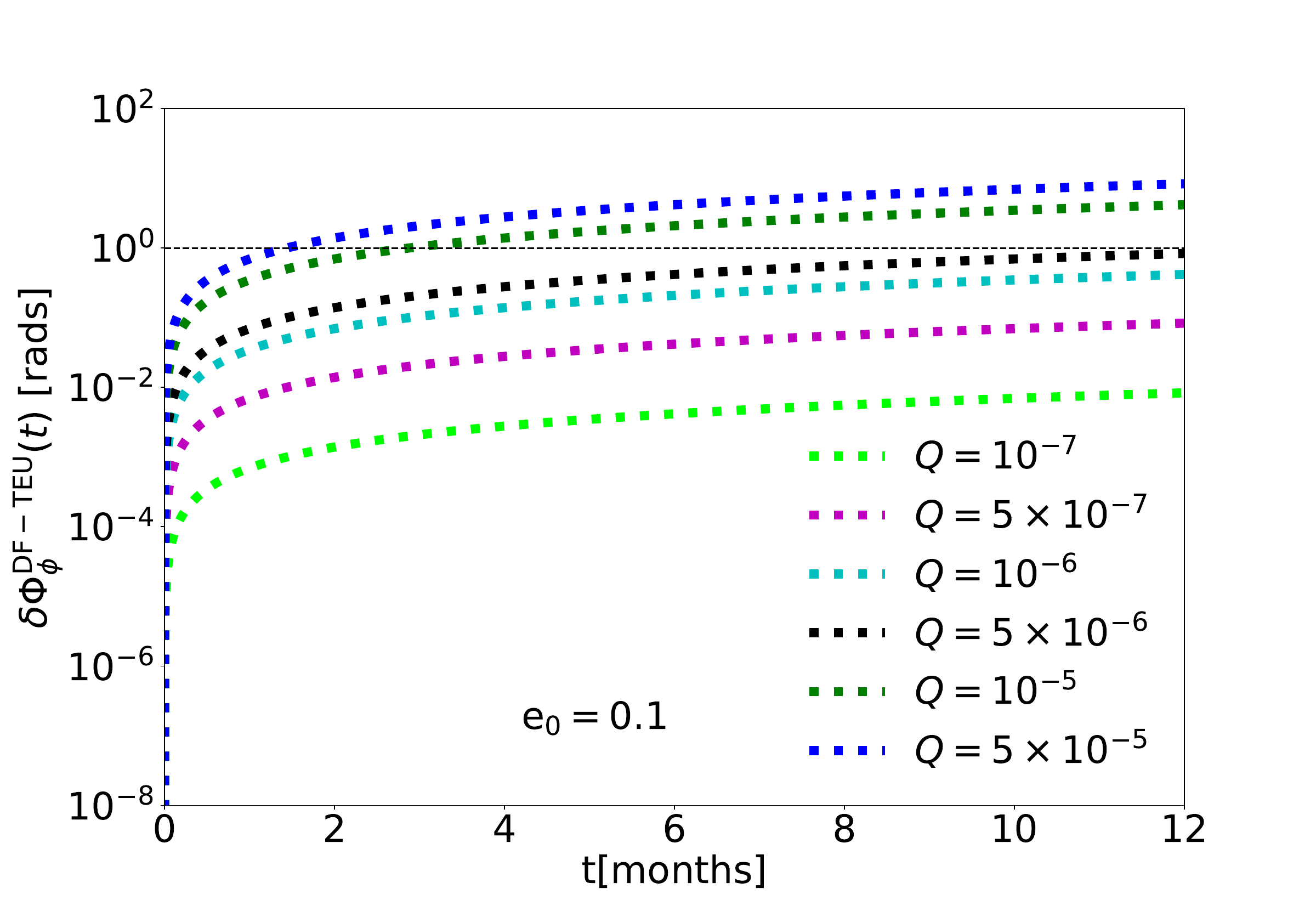}
\includegraphics[width=3.2in, height=2.2in]{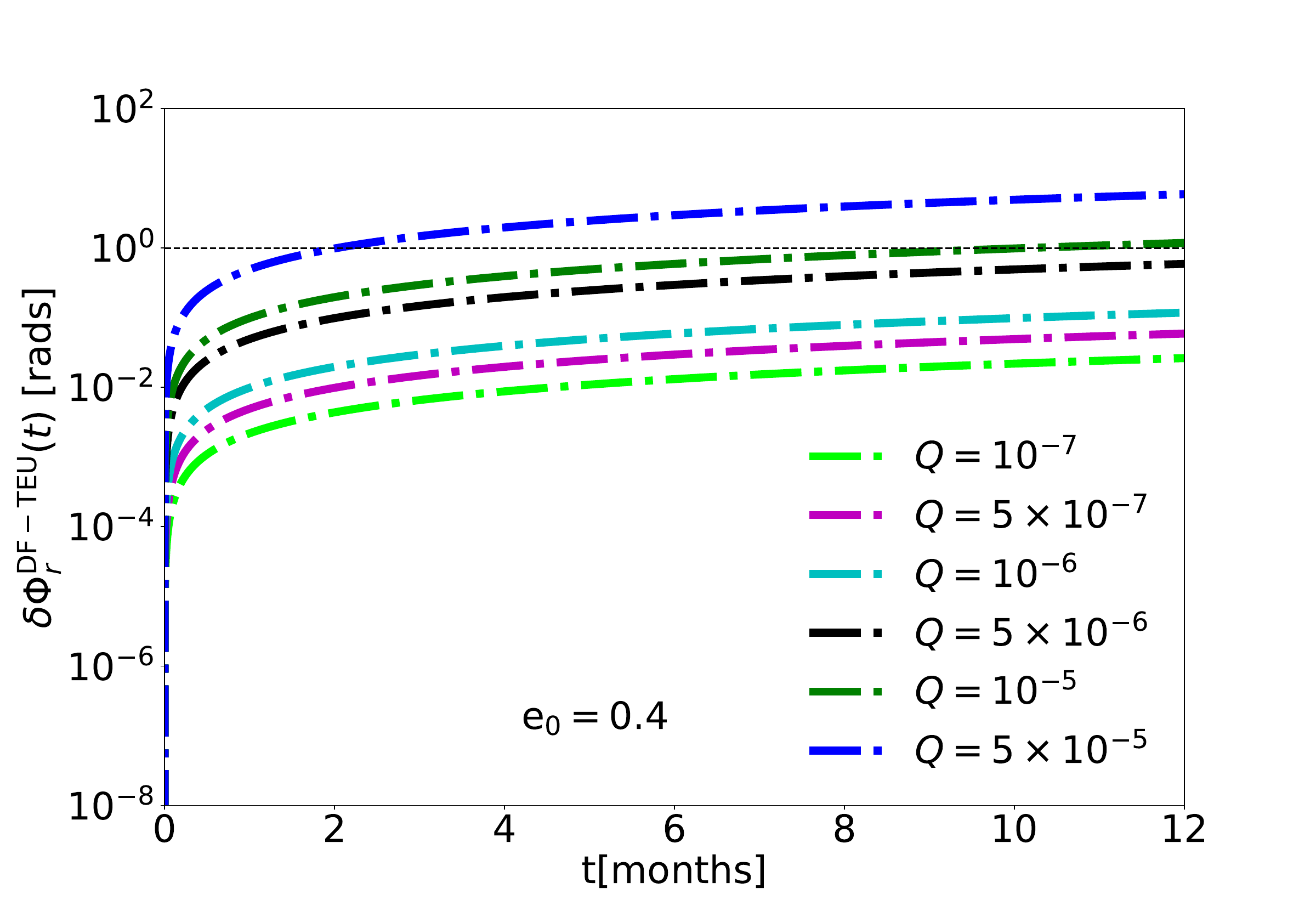}
\includegraphics[width=3.2in, height=2.2in]{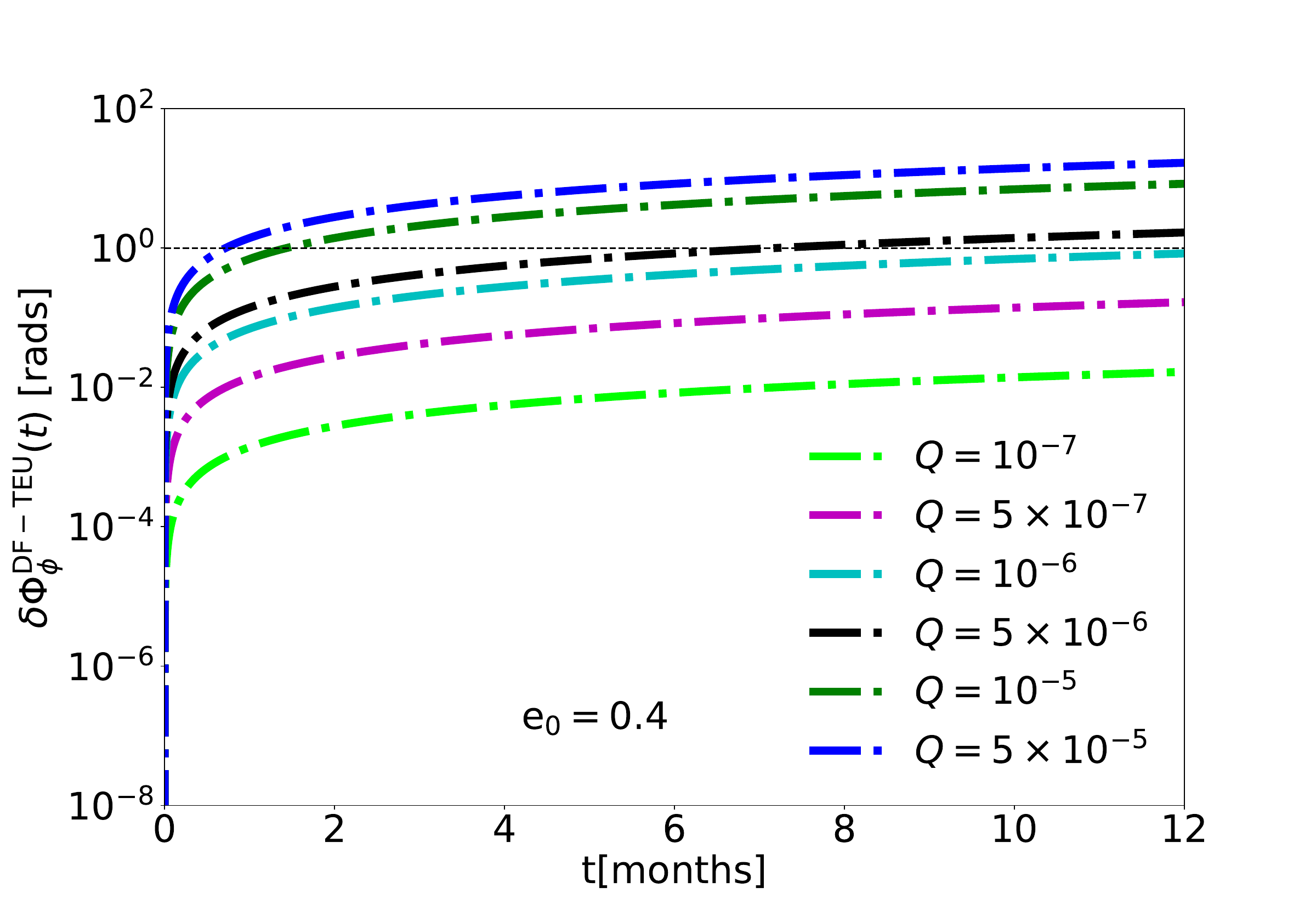}
\includegraphics[width=3.2in, height=2.2in]{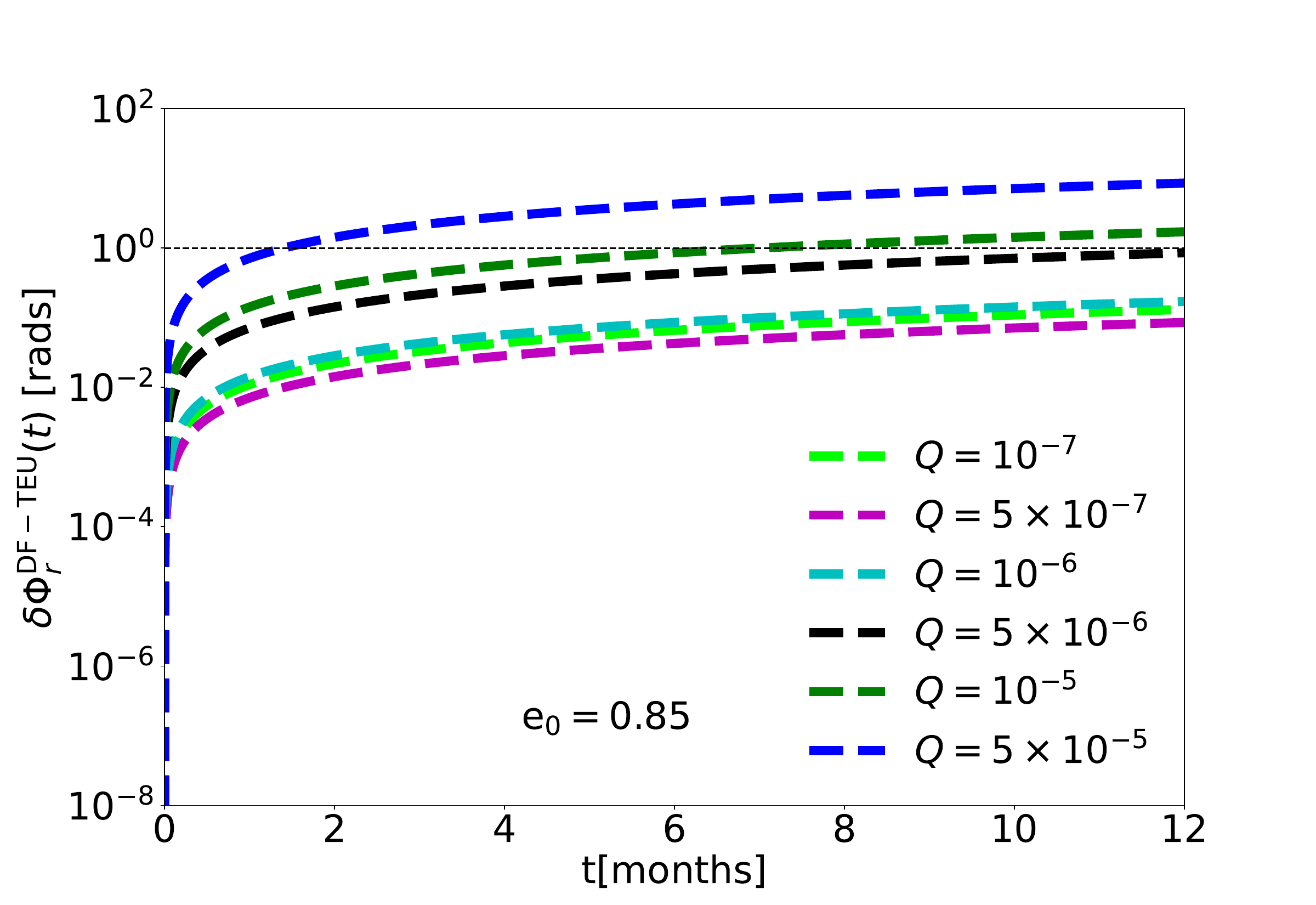}
\includegraphics[width=3.2in, height=2.2in]{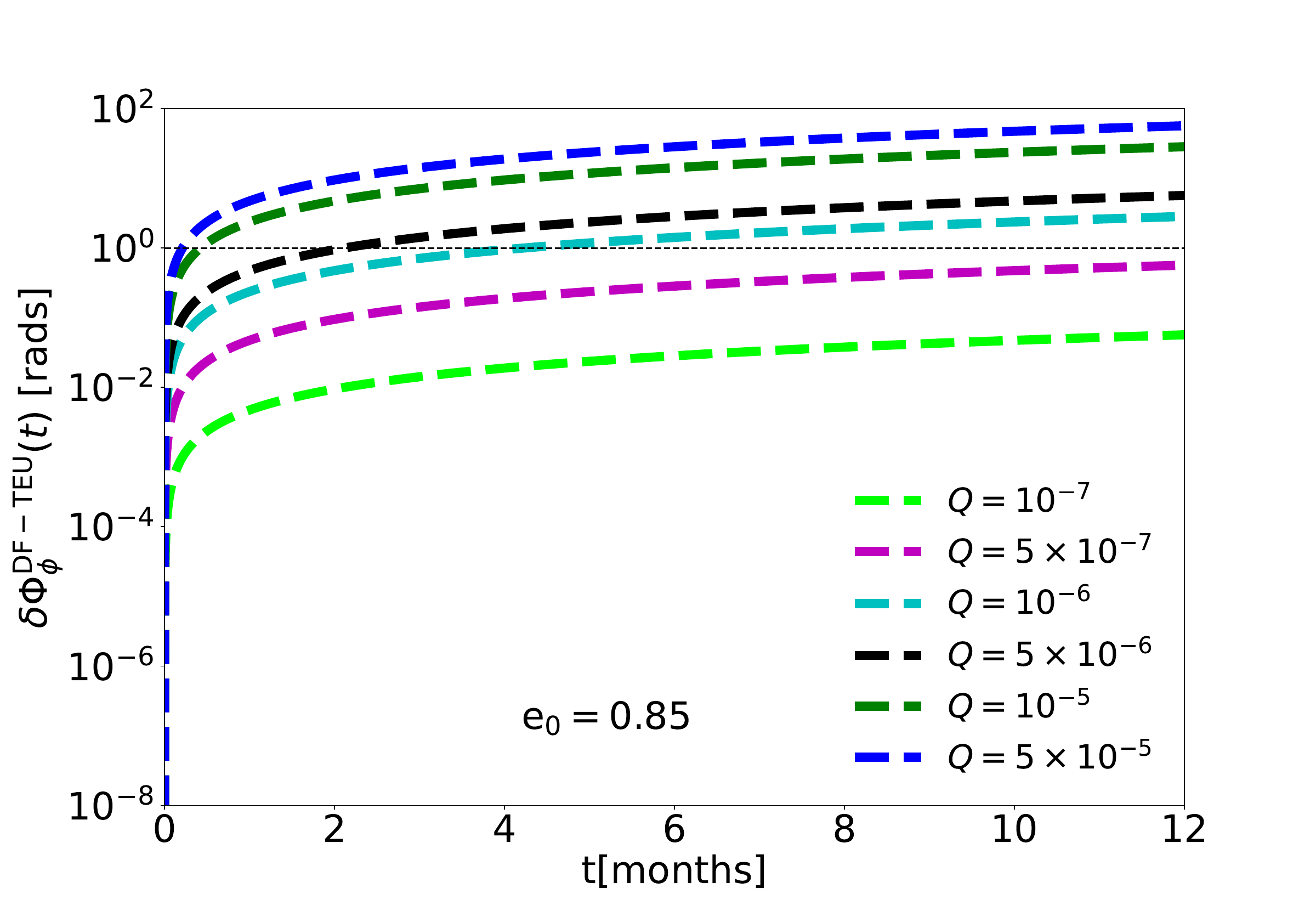}
\caption{The dephasing due to tidal charge is shown for initial eccentricity $e_0\in\{0.1,0.4,0.85\}$, in which the changing of evolution is derived by the fluxes from the standard Teukolsky equation of GR. The other intrinsic parameters are set as the mass of the central MBH $M=10^6M_\odot$, the initial orbital semi-latus rectum $p_0=12.0$ and tidal charge $Q\in\{10^{-7},5\times 10^{-7},10^{-6}, 5\times 10^{-6}, 10^{-5}, 5\times 10^{-5}\}$.} \label{fig:dephasing1}
\end{figure*}
\begin{align}\label{mismatch}
\mathcal{M}(h_{a}, h_{b}) = 1 - \mathcal{O}(h_{a}, h_{b})\,,
\end{align}
where overlap and inner product between two waveforms in the frequency domain are given by
\begin{equation}
\begin{aligned}\label{overlap}
\mathcal{O}(h_a, h_b) =& \frac{(h_{a}\vert h_{b})}{\sqrt{(h_{a}\vert h_{a})(h_{b}\vert h_{b})}}\,, \\
(h_a|h_b) =& 2 \int^{f_{high}}_{f_{low}} \frac{h_a^*(f)h_b(f)+h_a(f)h_b^*(f)}{S_n(f)}df\,,
\end{aligned}
\end{equation}
where $f_{low}=10^{-4}~\rm Hz$ and $f_{high}$ corresponds to the orbital frequency near the LSO. $S_{n}(f)$ is the power spectral density of LISA \cite{LISA:2017pwj}. The measure of mismatch reflects the degree of overlap between the two signals, thereby serving as an effective tool to assess the distinguishability of the tidal charge effect within the sensitivity range of detectors like LISA. It is important to restate that we compute mismatch for distinct values of tidal charge and eccentricity from two perspectives: one by implementing the DF Teukolsky equation and the second with the use of MTE. In both cases, with the obtained mismatches, we compare the constraint on the tidal charge and the importance of MTE in waveform modeling. The mismatch computed from the DF-Teukolsky equation is denoted as $\mathcal{M}^{\textup{DF-TEU}}$, and the mismatch obtained from the modified Teukolsky results is represented as $\mathcal{M}^{\textup{MTE}}$. From Eq. (\ref{mismatch}), a mismatch $\mathcal{M}=0$ implies identical waveforms with maximum overlap $\mathcal{O}=1$. In realistic settings, small deviations, such as those from tidal charge, can alter the waveform. To assess detectability, a threshold $\mathcal{M}\leq 1/(2\rho^{2})$ is used \cite{Flanagan:1997kp,Lindblom:2008cm}, where $\rho$ is the signal-to-noise ratio (SNR). For $\rho=20$, this gives $\mathcal{M}\approx 0.00125$ \cite{Babak:2017tow, Fan:2020zhy}, setting a benchmark for identifying whether such deviations from GR are observable and potentially indicative of extra dimensions.
\begin{figure*}[htb!]
\centering
\includegraphics[width=3.2in, height=2.2in]{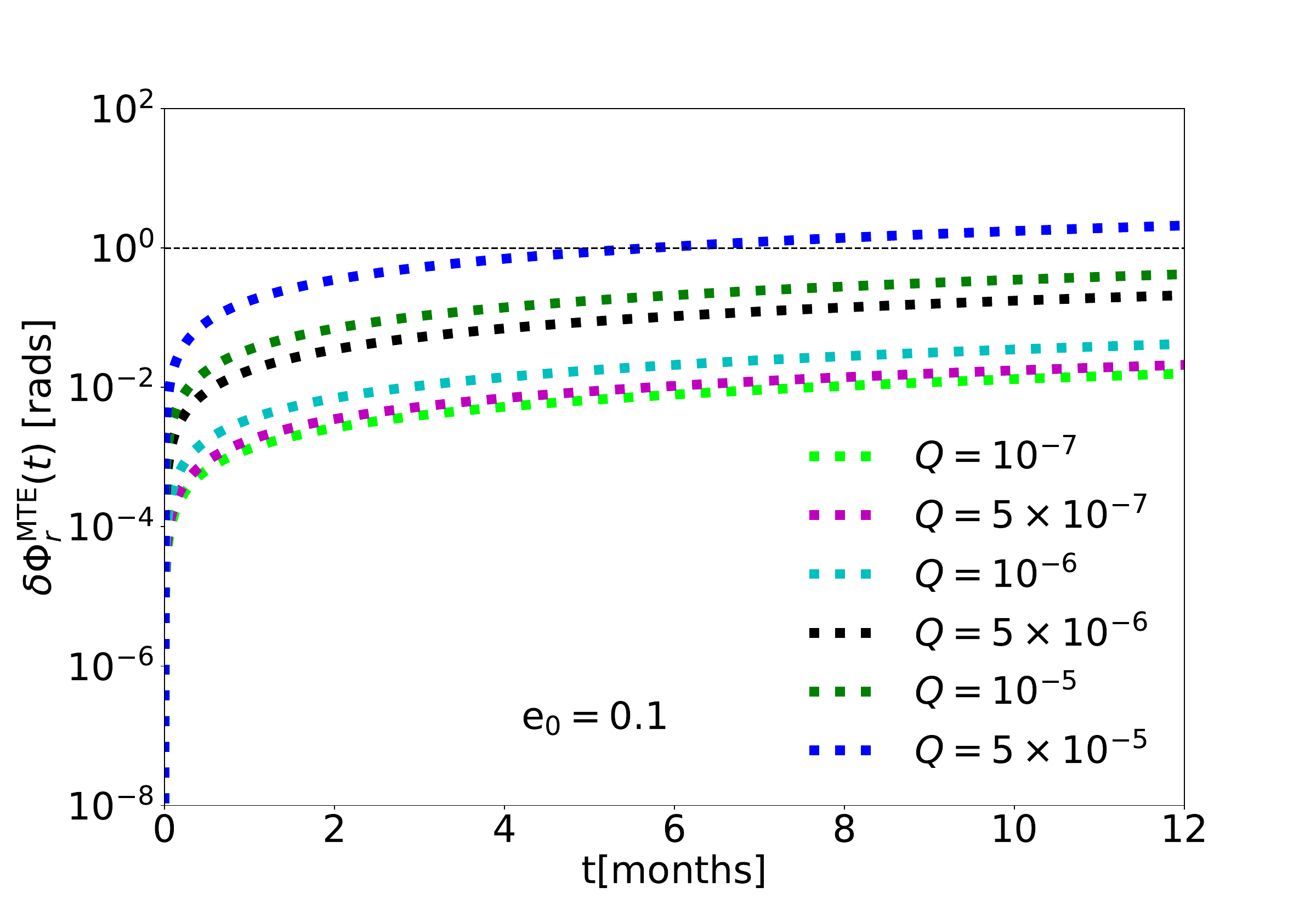}
\includegraphics[width=3.2in, height=2.2in]{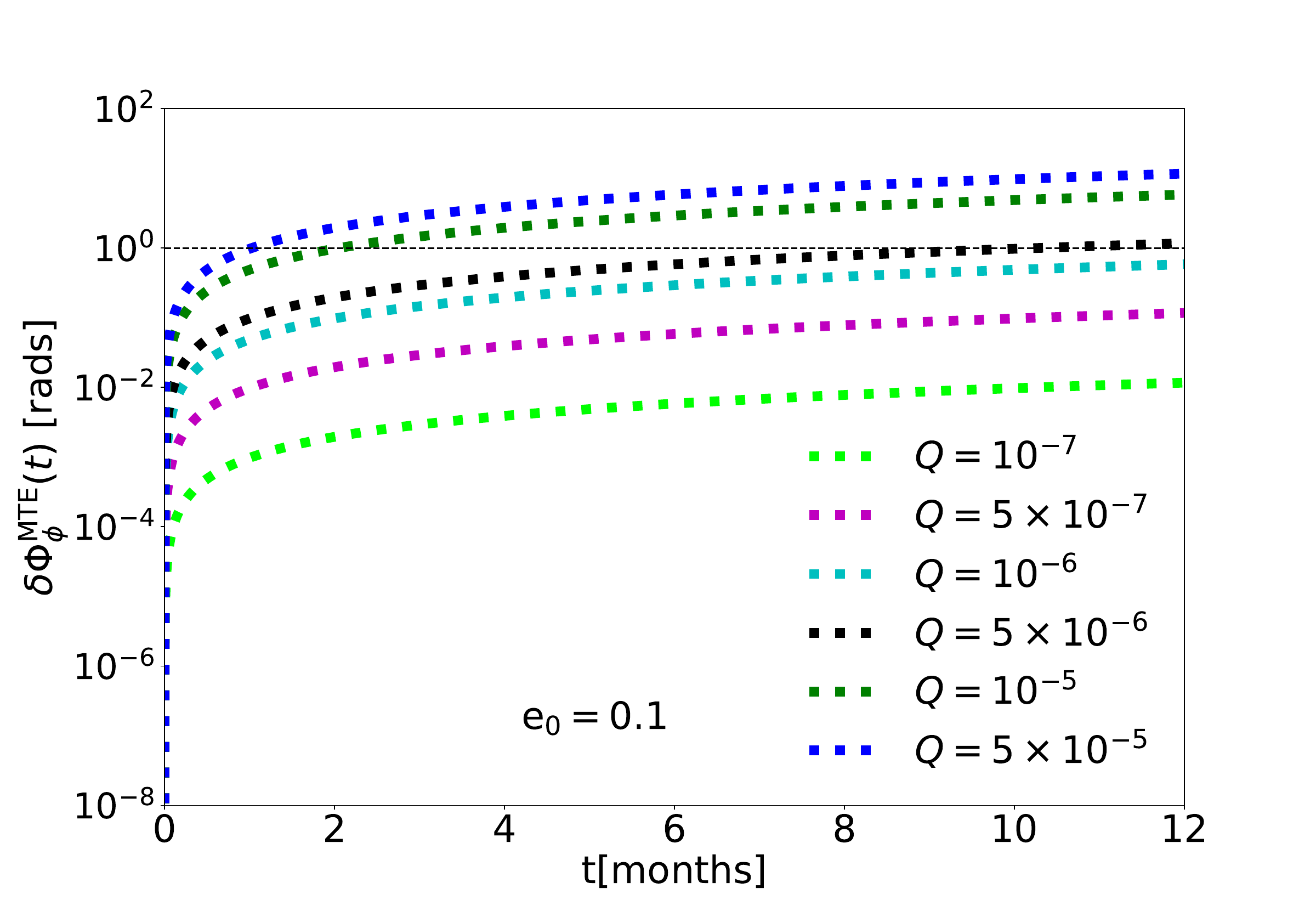}
\includegraphics[width=3.2in, height=2.2in]{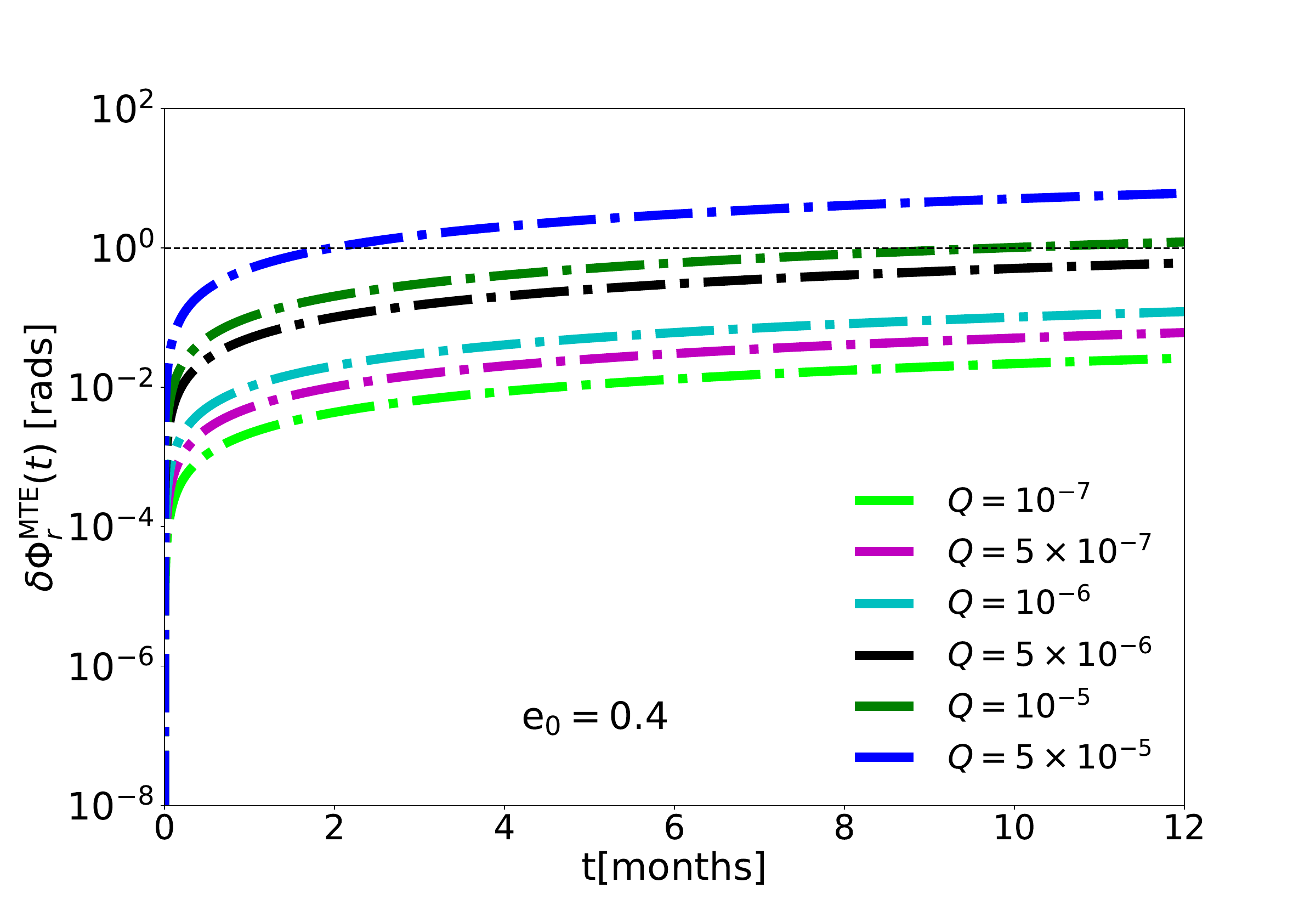}
\includegraphics[width=3.2in, height=2.2in]{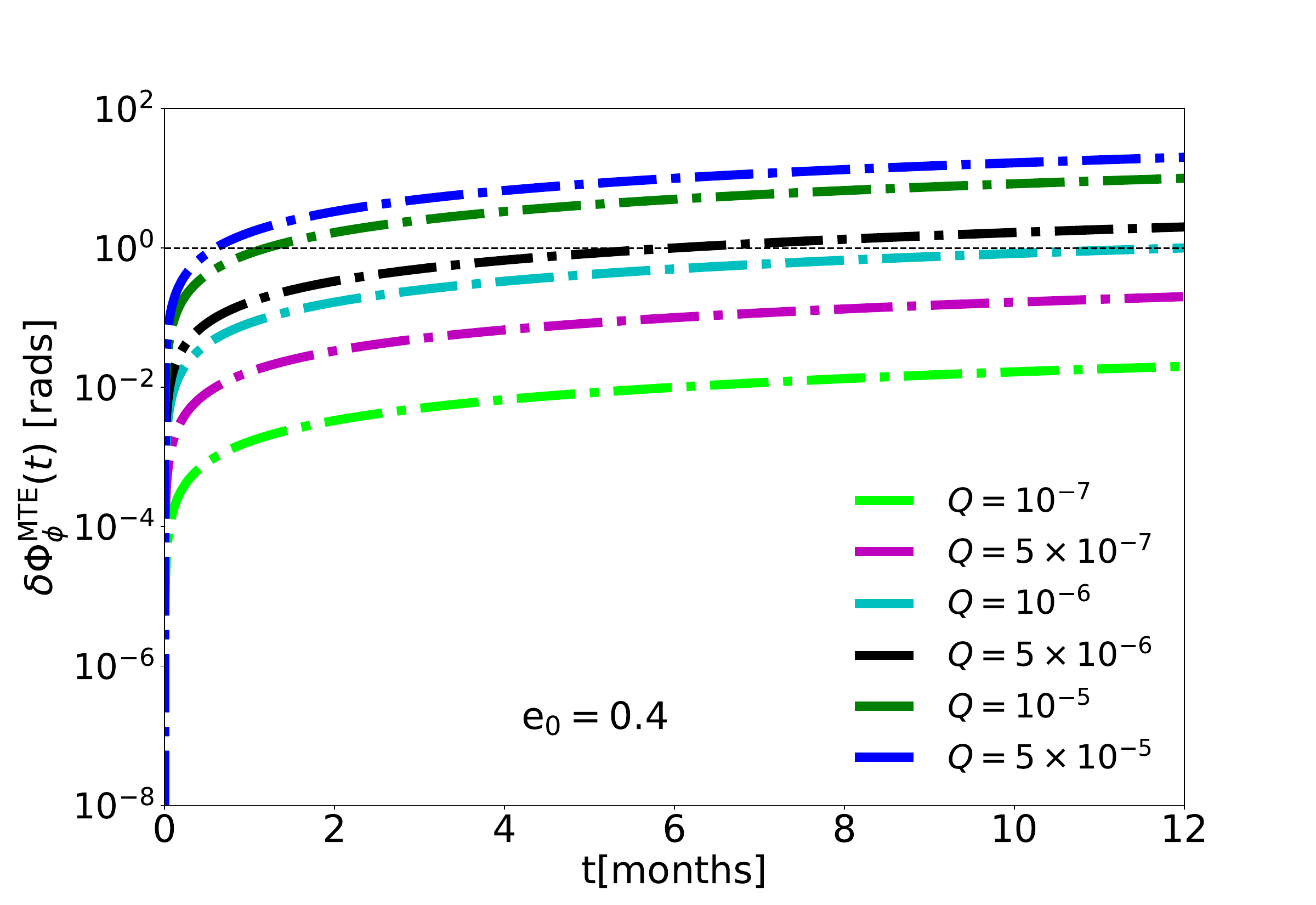}
\includegraphics[width=3.2in, height=2.2in]{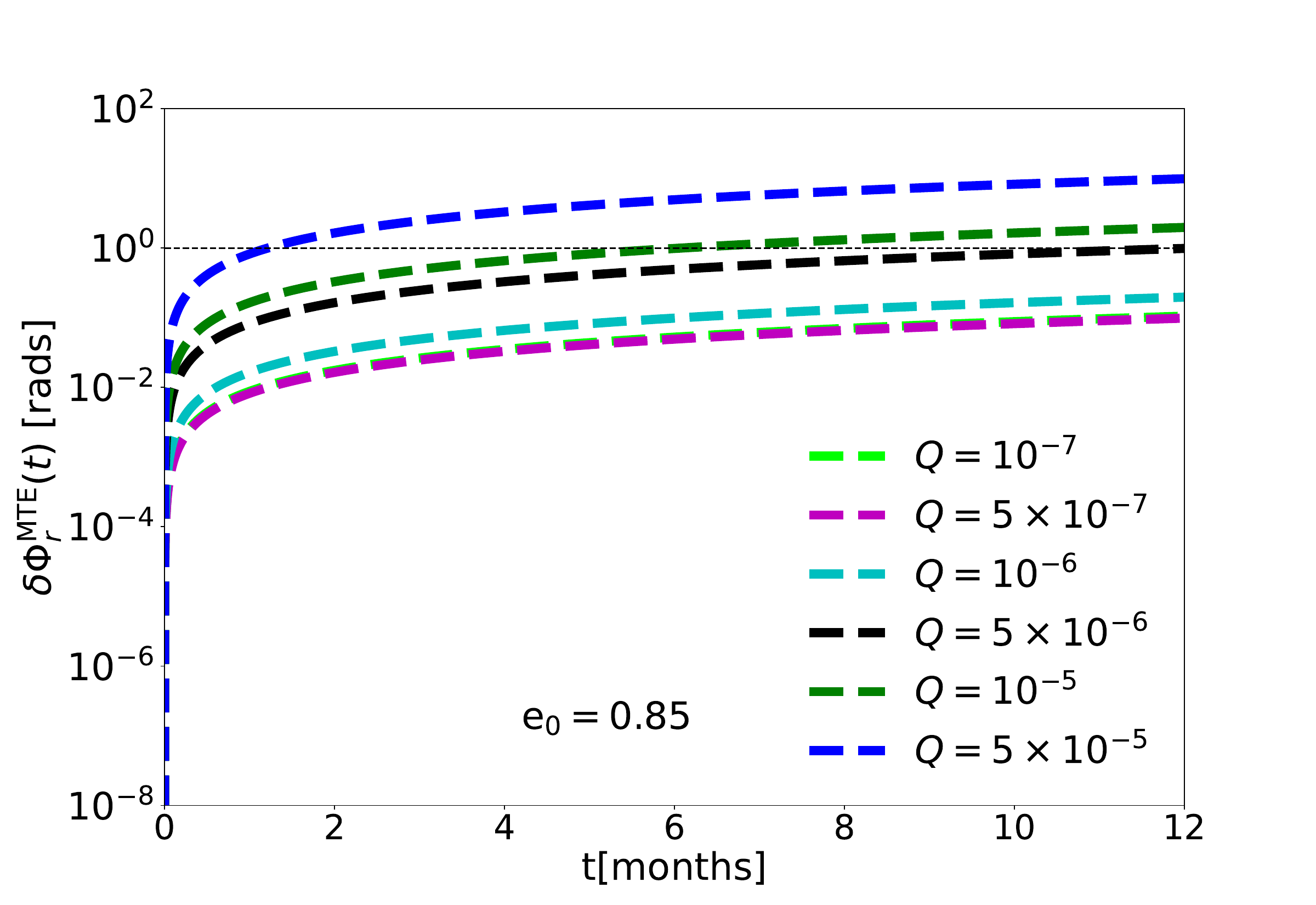}
\includegraphics[width=3.2in, height=2.2in]{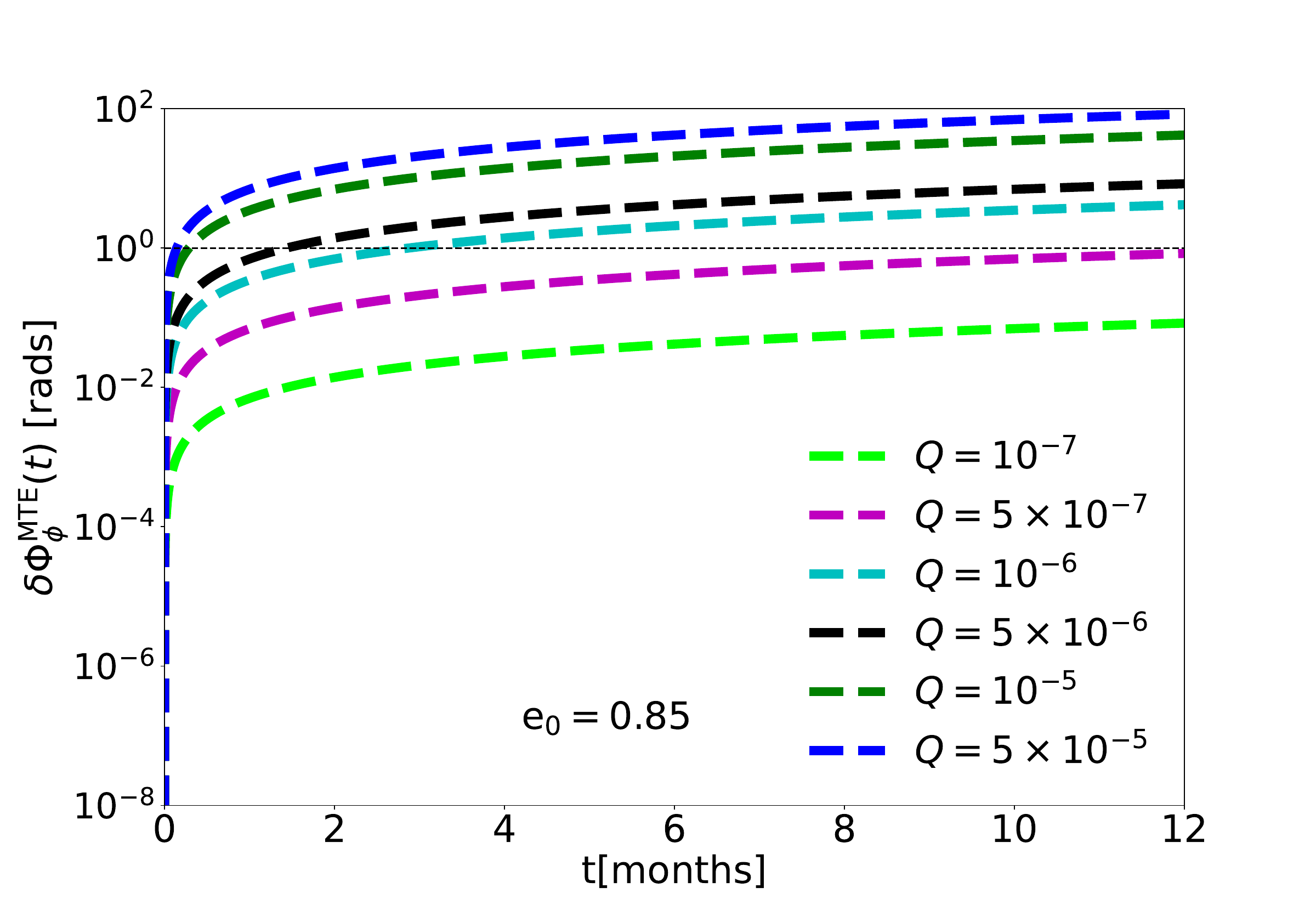}
\caption{The dephasing due to tidal charge as a function of observation time is plotted for initial eccentricity $e_0\in\{0.1,0.4,0.85\}$,
in which the changing of evolution is derived by the fluxes from MTE.
} \label{fig:dephasing2}
\end{figure*}



\section{Results} \label{wave}

\begin{figure*}[htb!]
\centering
\includegraphics[width=3.2in, height=2.2in]{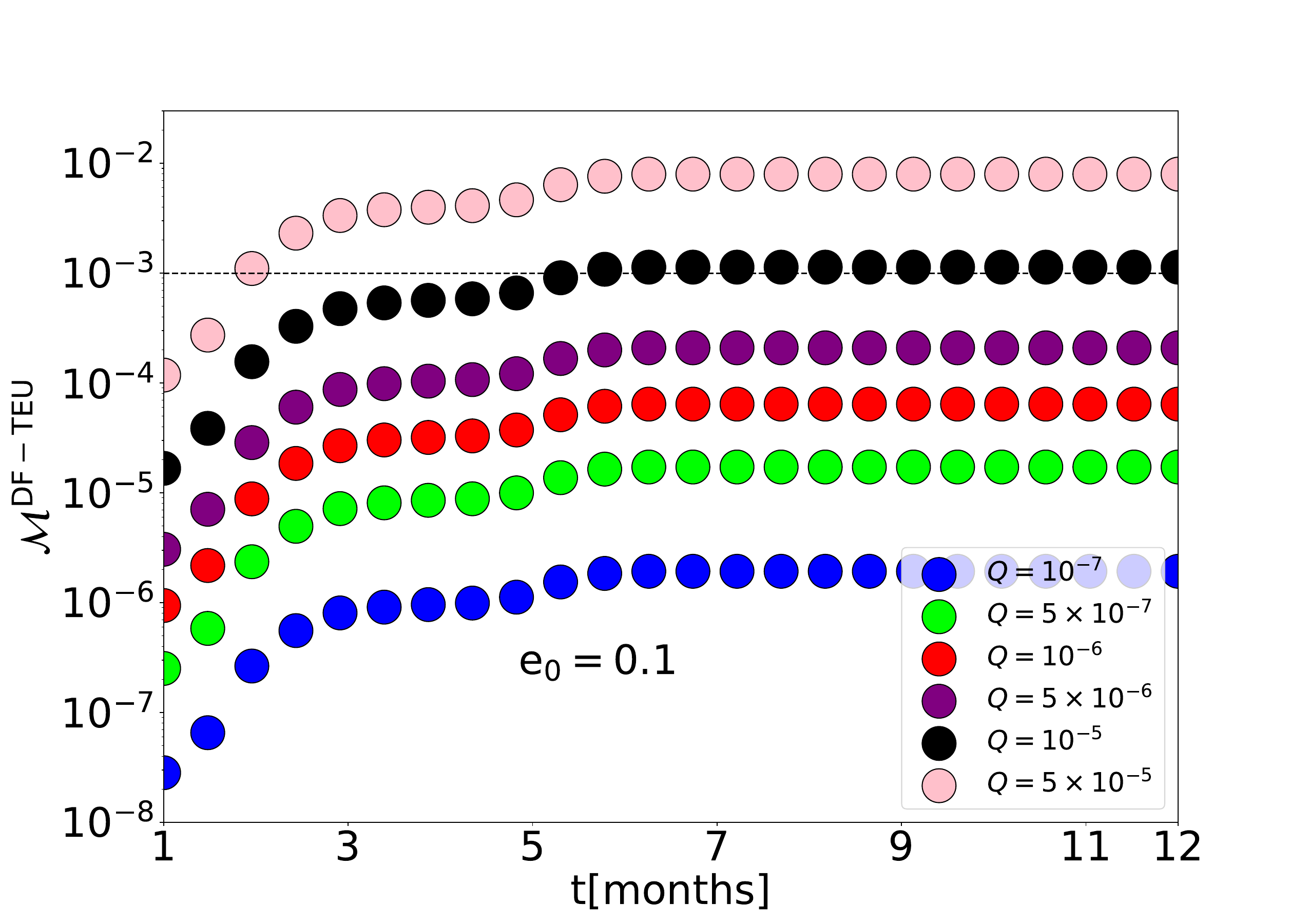}
\includegraphics[width=3.2in, height=2.2in]{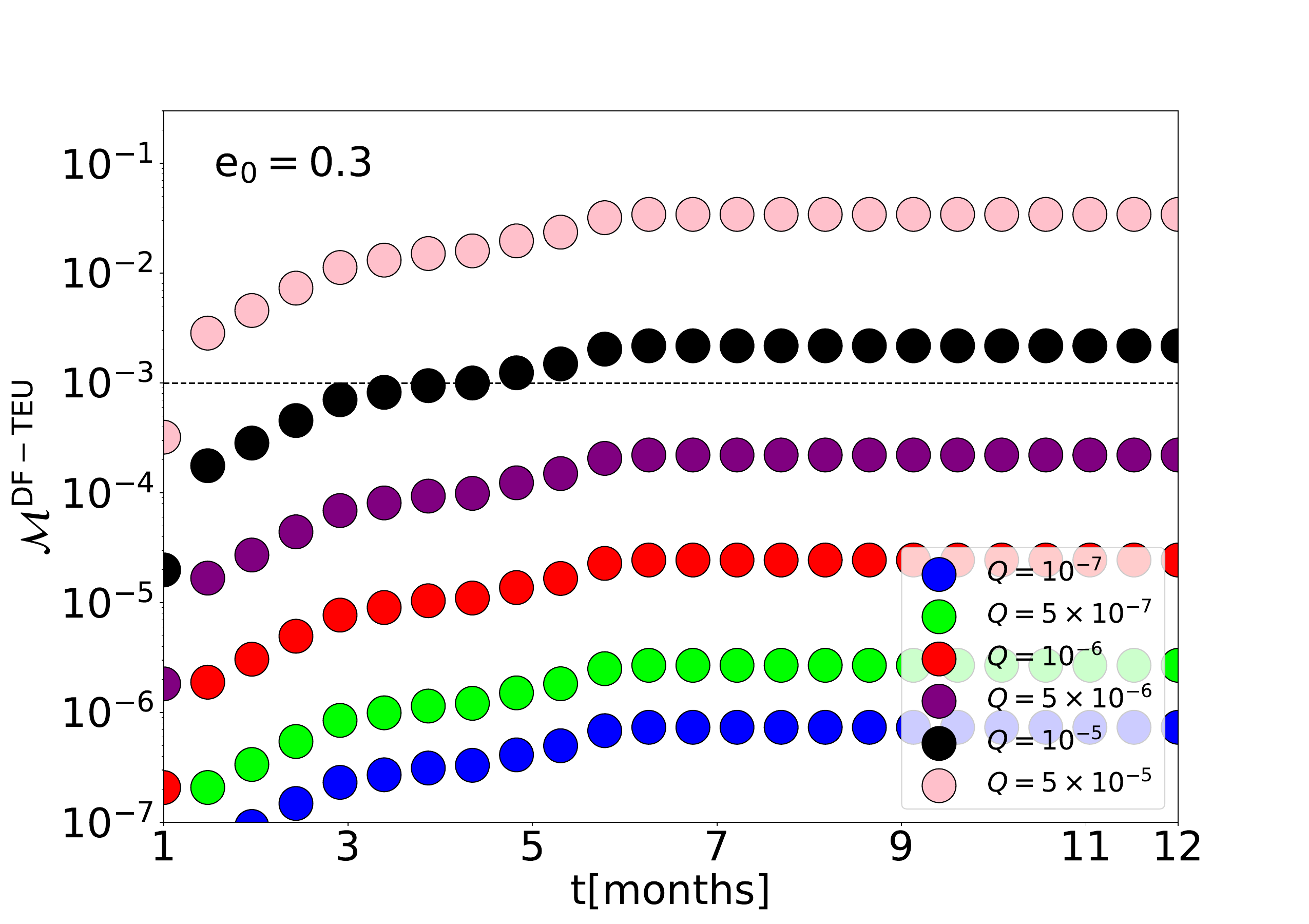}
\includegraphics[width=3.2in, height=2.2in]{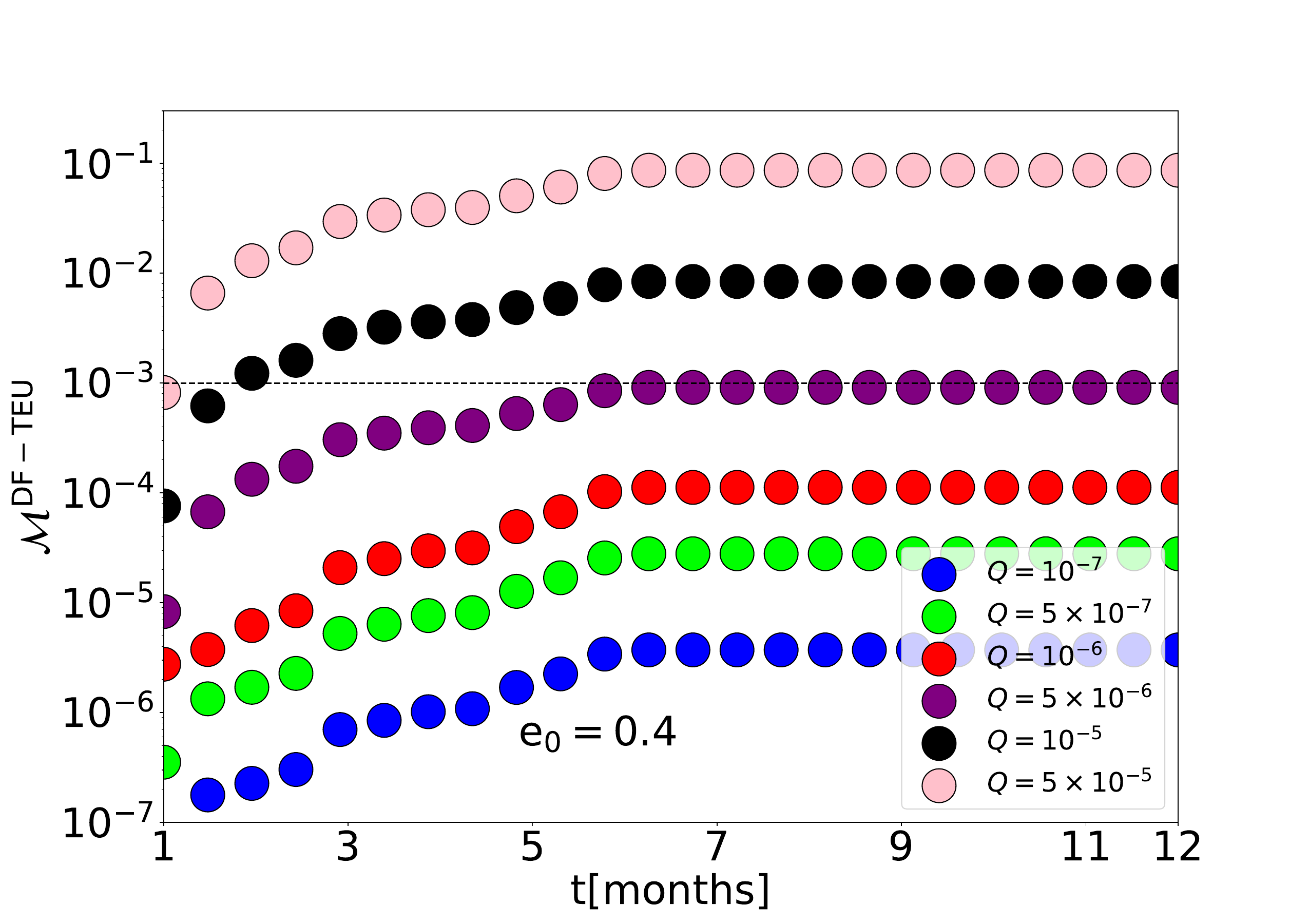}
\includegraphics[width=3.2in, height=2.2in]{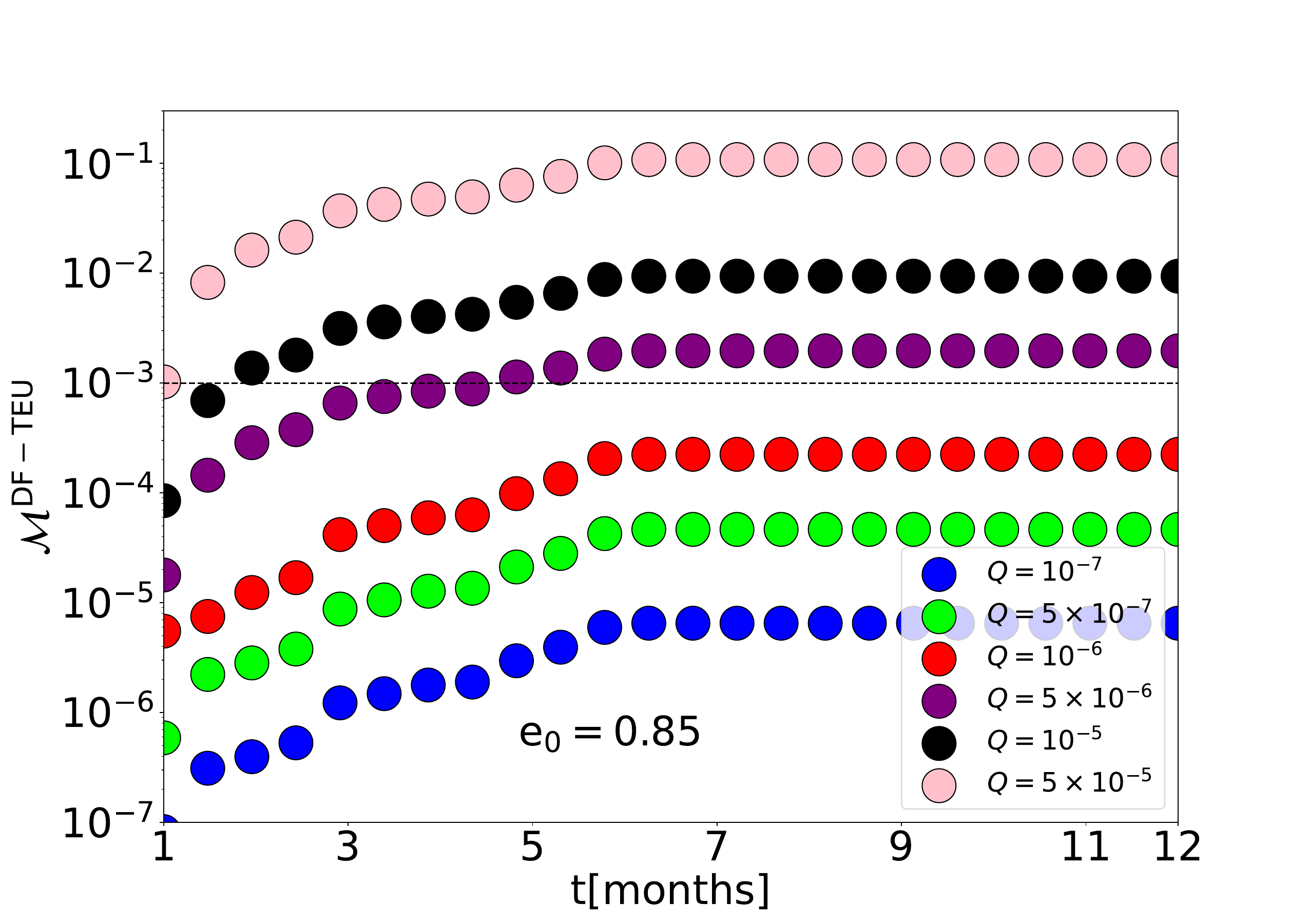}
\caption{Mismatch as a function of observation time for four initial orbital eccentricities $e_0 = (0.1,0.3,0.4,0.85)$ is plotted; the other parameters are $p_0=12.0$, $Q\in\{5\times 10^{-7}, 10^{-6}, 5\times 10^{-6}, 10^{-5}, 5\times10^{-5}, 10^{-4}\}$. Note that the mismatch is obtained from EMRI evolution using DF-Teukolsky fluxes in the presence and absence of $Q$.} \label{Fig:mismatch1}
\end{figure*}

Before taking a step ahead, we reiterate that in the braneworld case, neglecting bulk perturbations ($\delta E_{\mu\nu}=0$), the perturbation equation takes the same form as the Schwarzschild-Teukolsky equation with a modified $\Delta(r)$. We call that the \textit{DF-Teukolsky equation}, and reserving \textit{GR Teukolsky} for the vacuum Ricci-flat $Q=0$ case. However, the MTE description relaxes this assumption and relies on $\delta E_{\mu\nu}\neq 0$, incorporating the non-Ricci-flat nature of the spacetime. In this sense, in the results, we will see that Figs. (\ref{fig:dephasing1}) and (\ref{Fig:mismatch1}) correspond to the DF-Teukolsky setup, while Figs. (\ref{fig:dephasing2}) and (\ref{Fig:mismatch2}) employ the MTE prescription that includes the additional potential term $V_{s}(r)$ in Eq. (\ref{radequ}) (term proportional to $\varepsilon$). The comparison between these two cases allows us to quantify the impact of MTE corrections relative to the DF-Teukolsky results presented in Fig. (\ref{Fig:mismatch3}). Thus, our goal is to compare results obtained from the DF-Teukolsky equation ($\varepsilon = 0$) with modified $\Delta(r)$ and from the MTE equation ($\varepsilon = 1$) with modified $\Delta(r)$, and to highlight the significance of the MTE framework for the precision computation of GW observables in theories beyond GR. Note that in both approaches, the reference waveform is obtained from the standard Teukolsky equation for the Schwarzschild spacetime, where $Q = 0$.
\begin{figure*}[htb!]
\centering
\includegraphics[width=3.2in, height=2.2in]{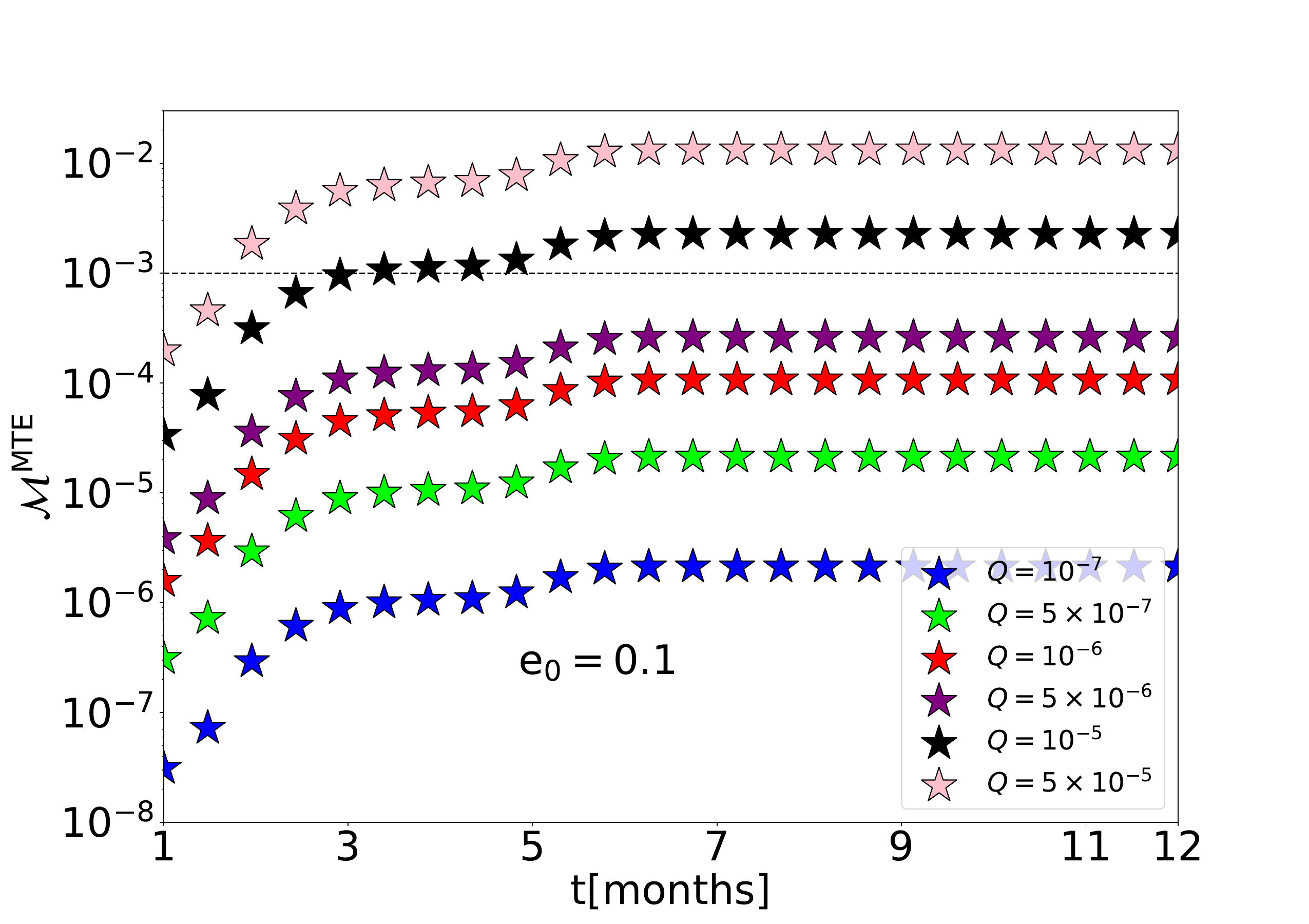}
\includegraphics[width=3.2in, height=2.2in]{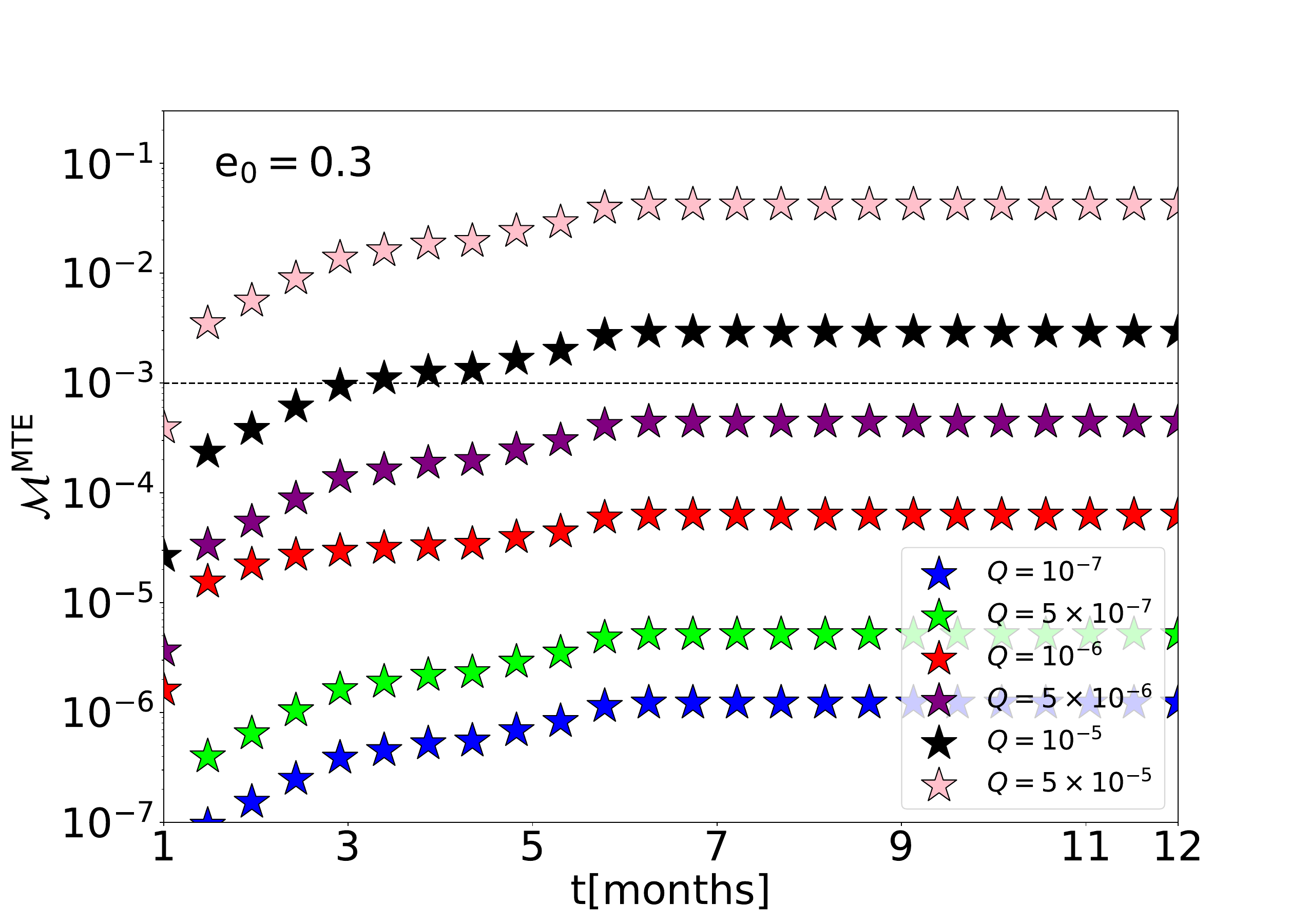}
\includegraphics[width=3.2in, height=2.2in]{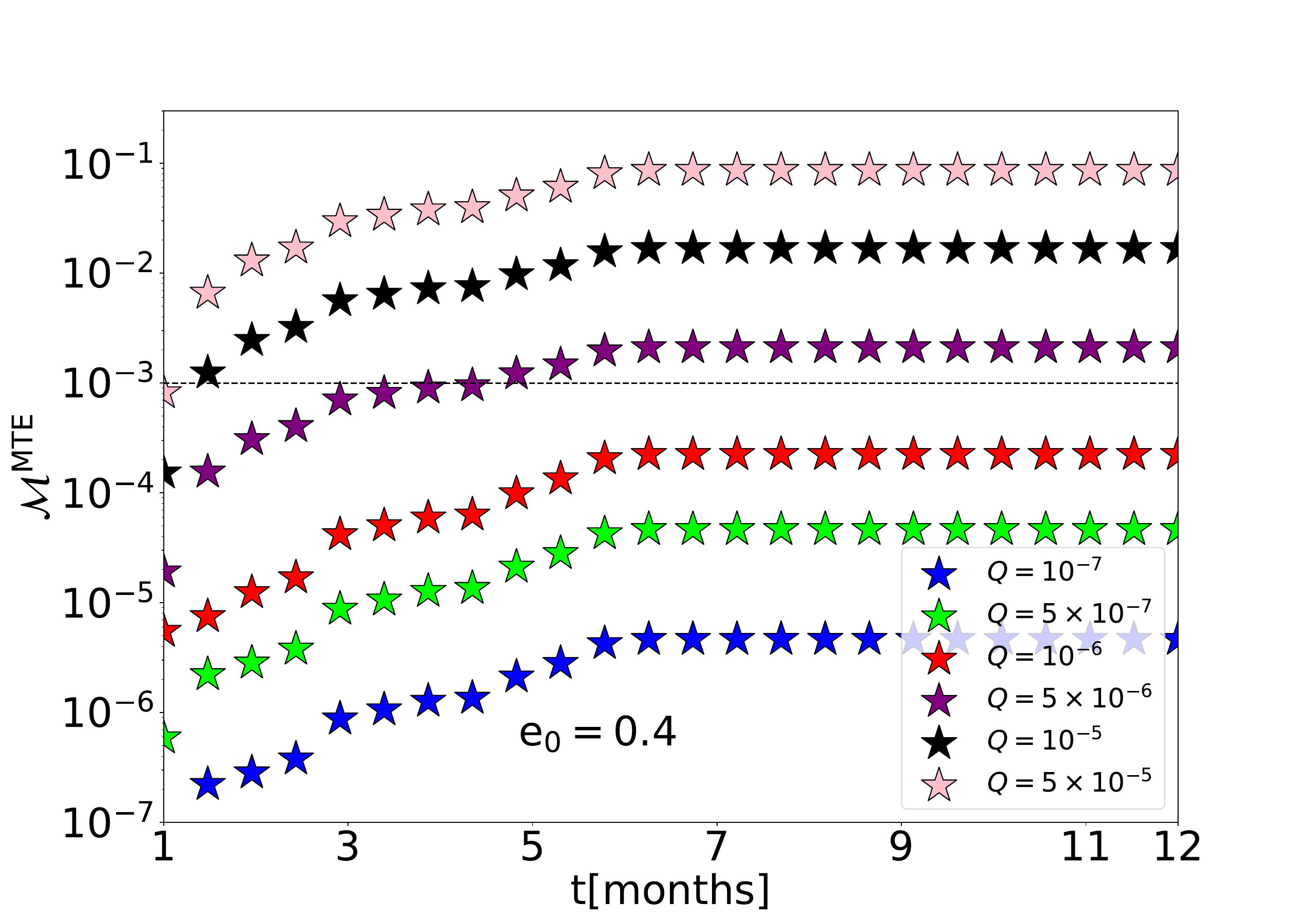}
\includegraphics[width=3.2in, height=2.2in]{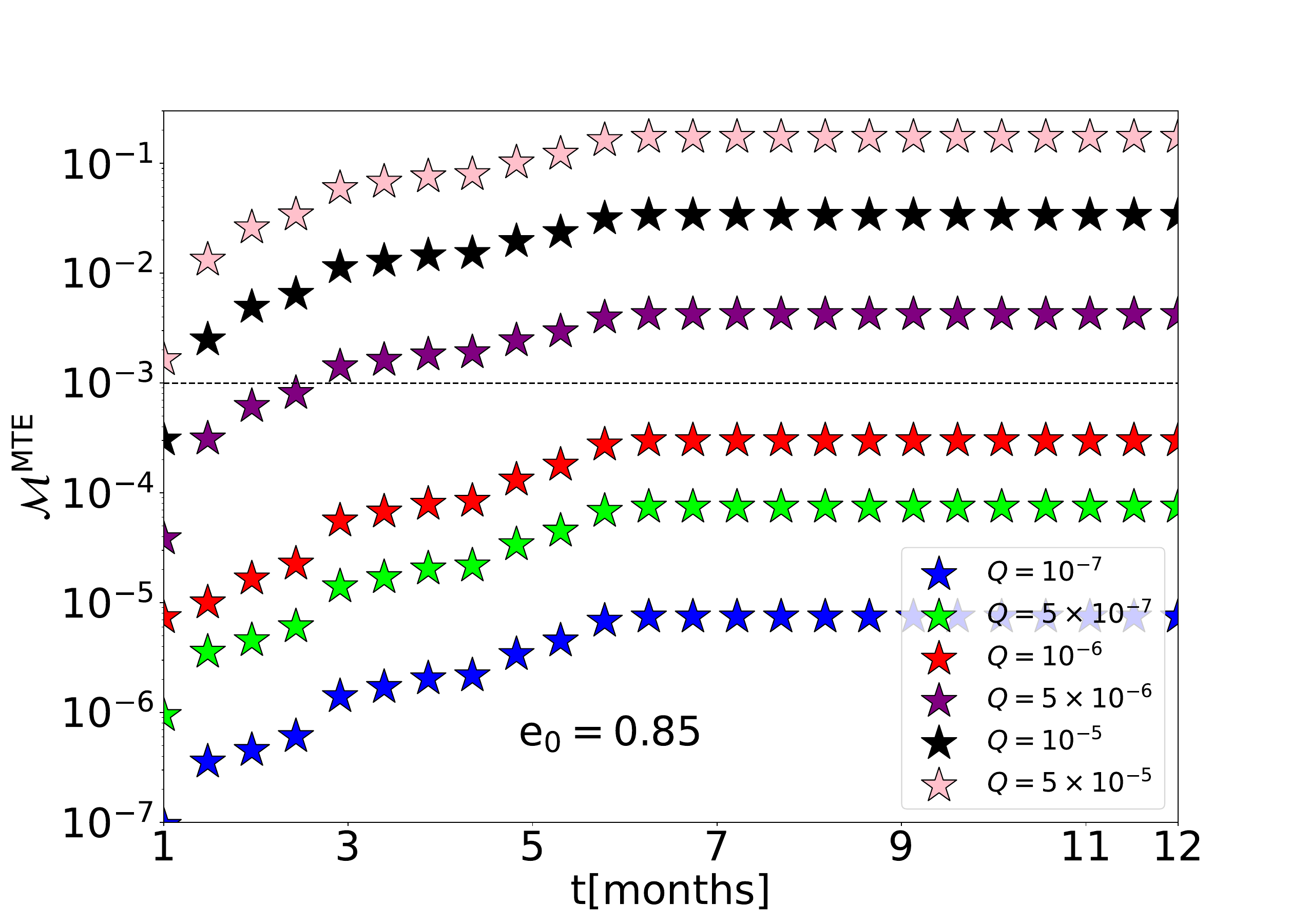}
\caption{Mismatch as a function of observation time for four initial orbital eccentricities $e_0 = (0.1,0.3,0.4, 0.85)$ is plotted, the other parameters are $p_0=12.0$, $Q\in\{5\times 10^{-7}, 10^{-6}, 5\times 10^{-6}, 10^{-5}, 5\times10^{-5}, 10^{-4}\}$. Note that $h^{\rm MTE}$ is the waveform from EMRIs evolution using modified Teukolsky fluxes and $h^{\rm GR}$ denotes to the waveform obtained from the standard GR fluxes.} \label{Fig:mismatch2}
\end{figure*}
Prior to discussing the detailed results, we provide a brief overview of the findings and their importance. We emphasize, following the dephasing and mismatch plots for both the DF-Teukolsky equation and the MTE cases, that the curves cross the detection threshold\textemdash marked by the horizontal line\textemdash around $\sim 1$ rad in the dephasing plots and $\sim 10^{-3}$ in the mismatch plots, occurring at approximately the same value of the tidal charge $Q$, as also reflected in Table~(\ref{tab1}). This observation is reiterated here to underline the consistency between the two approaches and to reinforce the robustness of the result. Furthermore, while the upper bound (the bound below which LISA might not detect the corresponding dephasing/mismatch) on $Q$ remains nearly unchanged, both the dephasing and mismatch increase with orbital eccentricity ($e_0$). Importantly, the difference in mismatch between the DF-Teukolsky and MTE cases grows with increasing eccentricity, clearly demonstrating the added sensitivity that this MTE formulation brings to this study. This behavior underscores the significance of the present analysis, illustrating how eccentricity amplifies observable deviations in EMRI signals within the braneworld framework.

Now, we begin with describing the results obtained for the prospects of determining the existence of extra dimensions through GW observations with the comparison between the DF-Teukolsky and MTE approaches, as well as the need of the MTE for more accurate and pronounced results. Before describing the results, we recall the following definitions: the results/plots written with `DF-TEU' in superscript denote the (approximate) analysis based on the Dudley-Finley Teukolsky equation, whereas the results with `MTE' in superscript define the analysis based on the modified Teukolsky framework as discussed in Section~(\ref{perturbation}). Note that in the results derived, we set the mass-ratio $q=10^{-5}$ and the secondary starts the inspiral from $p_{0}=12$. Let us now discuss the key results in detail.
\begin{alphalist}
\item We start with the dephasing results, whose detection criterion is $\sim 1$ rad. In Fig. (\ref{fig:dephasing1}), with the use of fluxes and evolution obtained under the DF-Teukolsky implementation, we provide radial and azimuthal dephasings over the observation period of one year for distinct values of ($Q, e_0$). We can easily see that the azimuthal dephasing ($\delta\Phi^{\textup{DF-TEU}}_{\phi}$) is larger than the radial dephasing ($\delta\Phi^{\textup{DF-TEU}}_{r}$), consistent with the existing literature. As we increase the eccentricity, the dephasing increases; correspondingly, the constraint on the tidal charge also changes. For instance, we notice that for azimuthal dephasing, the constraint on $Q$ changes from $10^{-5}$ to $10^{-6}$ for $e_0=0.1$ to $e_0=0.85$, respectively. A similar case can be seen with radial dephasing.

\item In Fig.~(\ref{fig:dephasing2}), we provide the dephasing using the fluxes and orbital evolution obtained through MTE. Similar to Fig. (\ref{fig:dephasing1}), we plot the radial and azimuthal dephasings for distinct values of ($Q, e_0$). We also notice that as we increase eccentricity, the corresponding dephasing also increases. The inclusion of the correction coming from the MTE introduces changes in the dephasings. The constraint on $Q$ changes from $\sim 5\times 10^{-6}$ to $\sim 10^{-6}$ for $e_0=0.1$ to $e_0=0.85$. It is apparent from Fig. (\ref{fig:dephasing1}) and (\ref{fig:dephasing2}) that the dephasing that comes from the MTE approach is more than the one obtained from the DF-Teukolsky equation.
\end{alphalist}

Further, in Fig. (\ref{Fig:mismatch1}) and Fig. (\ref{Fig:mismatch2}), we present a mismatch-based analysis comparing waveforms generated using the DF-Teukolsky formalism ($\mathcal{M}^{\textup{DF-TEU}}$) and those obtained via the MTE approach ($\mathcal{M}^{\textup{MTE}}$). We explore various combinations of eccentricity ($e_0$) and tidal charge ($Q$), applying a detection threshold of $\mathcal{M} \approx 0.00125$, corresponding to an SNR of 20, to assess detectability.

\begin{alphalist}
\item The results show that both higher eccentricities and larger tidal charges lead to increased mismatch values. In the case of DF-Teukolsky waveforms (Fig. \ref{Fig:mismatch1}), a tidal charge of $\sim 10^{-5}$ reaches the detection threshold for $e_0 = 0.1$, while the constraint tightens to $\sim 5 \times 10^{-6}$ for $e_0 = 0.85$.

\item A similar trend is evident in the MTE-based results (Fig. \ref{Fig:mismatch2}); however, the mismatch values in the MTE case are higher than those coming from the DF-Teukolsky for higher values of eccentricity. For example, at $e_0 = 0.85$, the $\mathcal{M}^{\textup{MTE}}$ curve crosses the detection threshold earlier than $\mathcal{M}^{\textup{DF-TEU}}$\,.
This enhanced sensitivity underscores the utility of using the MTE framework for precision tests of GR.
\end{alphalist}
\begin{figure*}[t!]
\centering
\includegraphics[width=3.17in, height=2.2in]{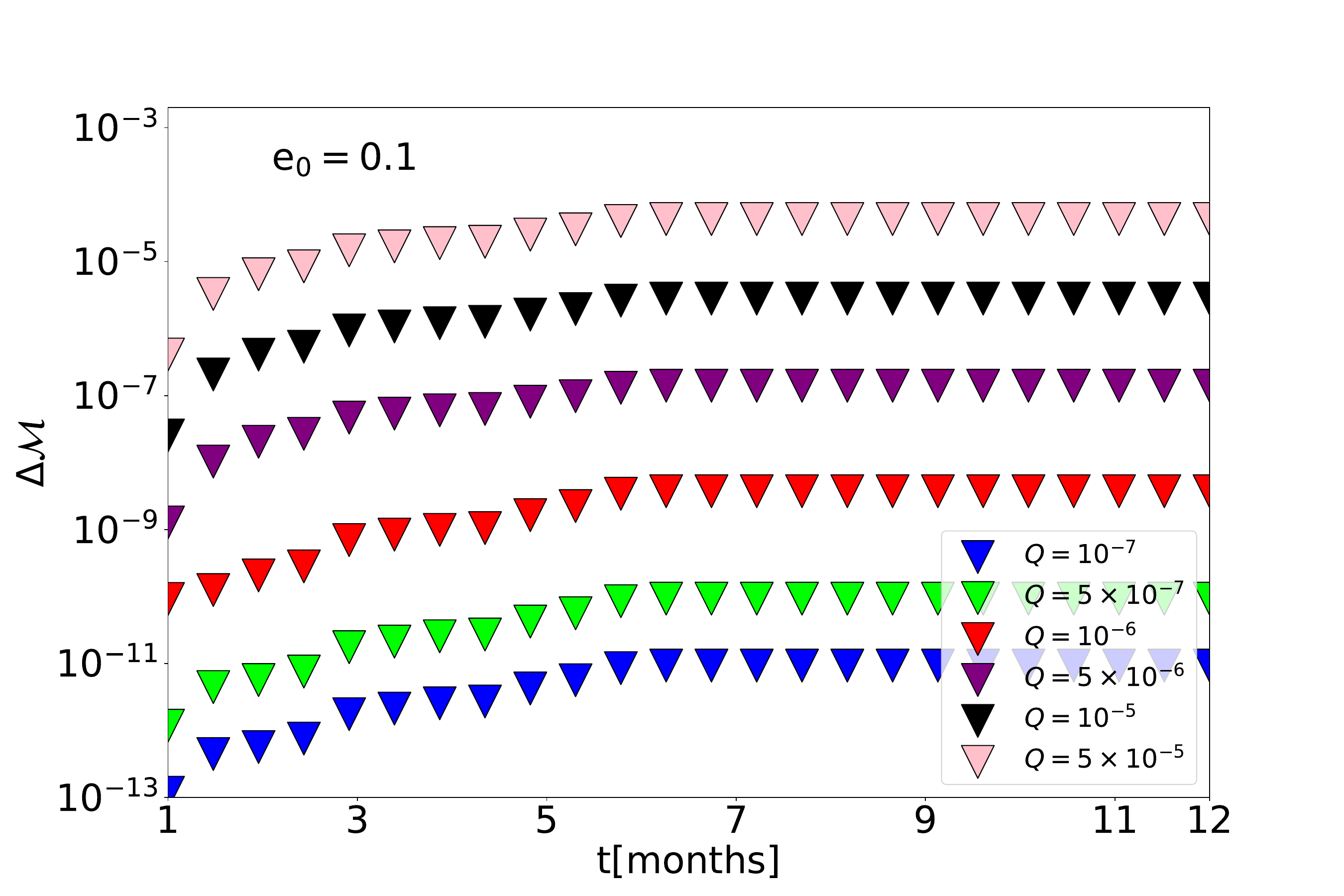}
\includegraphics[width=3.17in, height=2.2in]{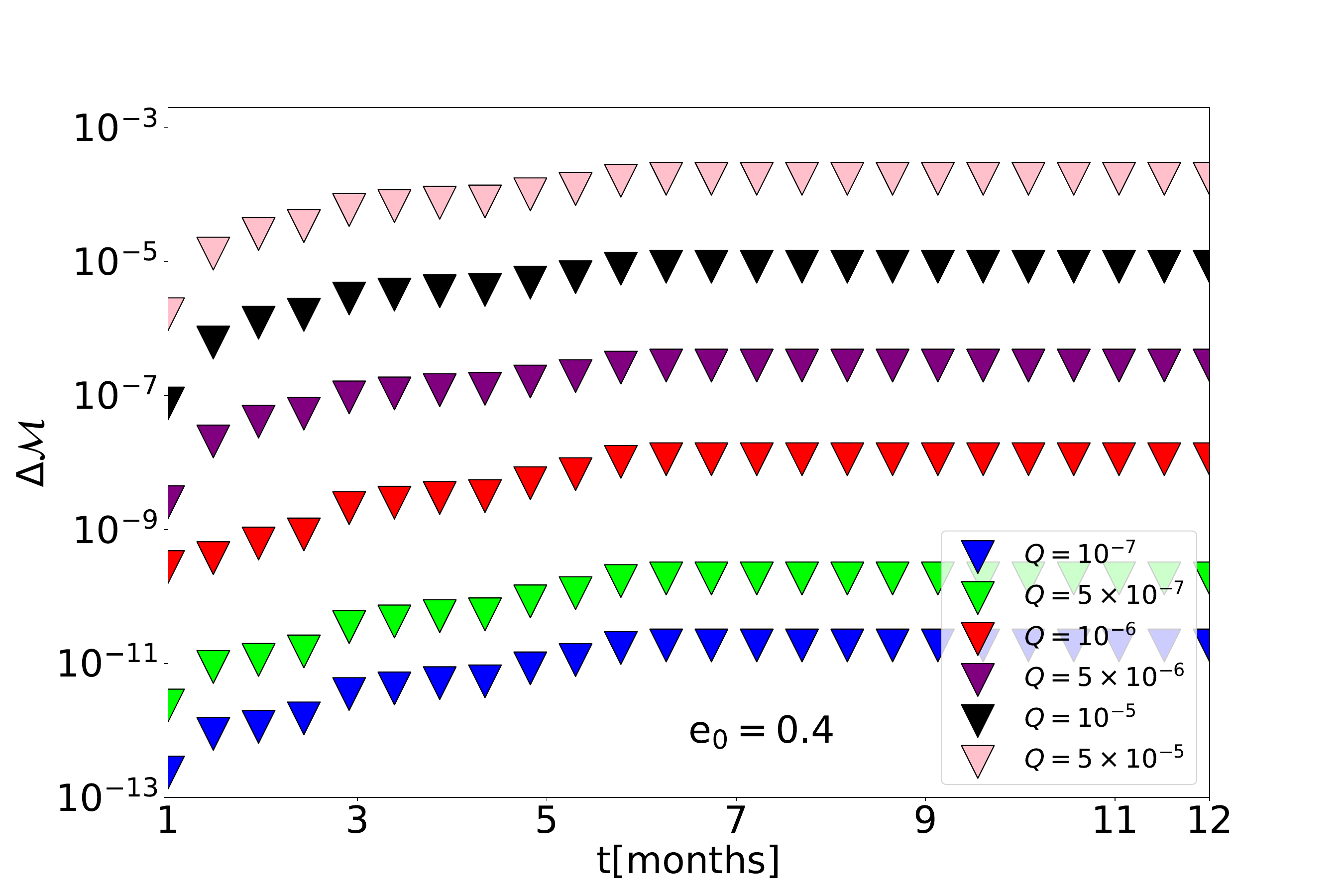}
\includegraphics[width=3.17in, height=2.2in]{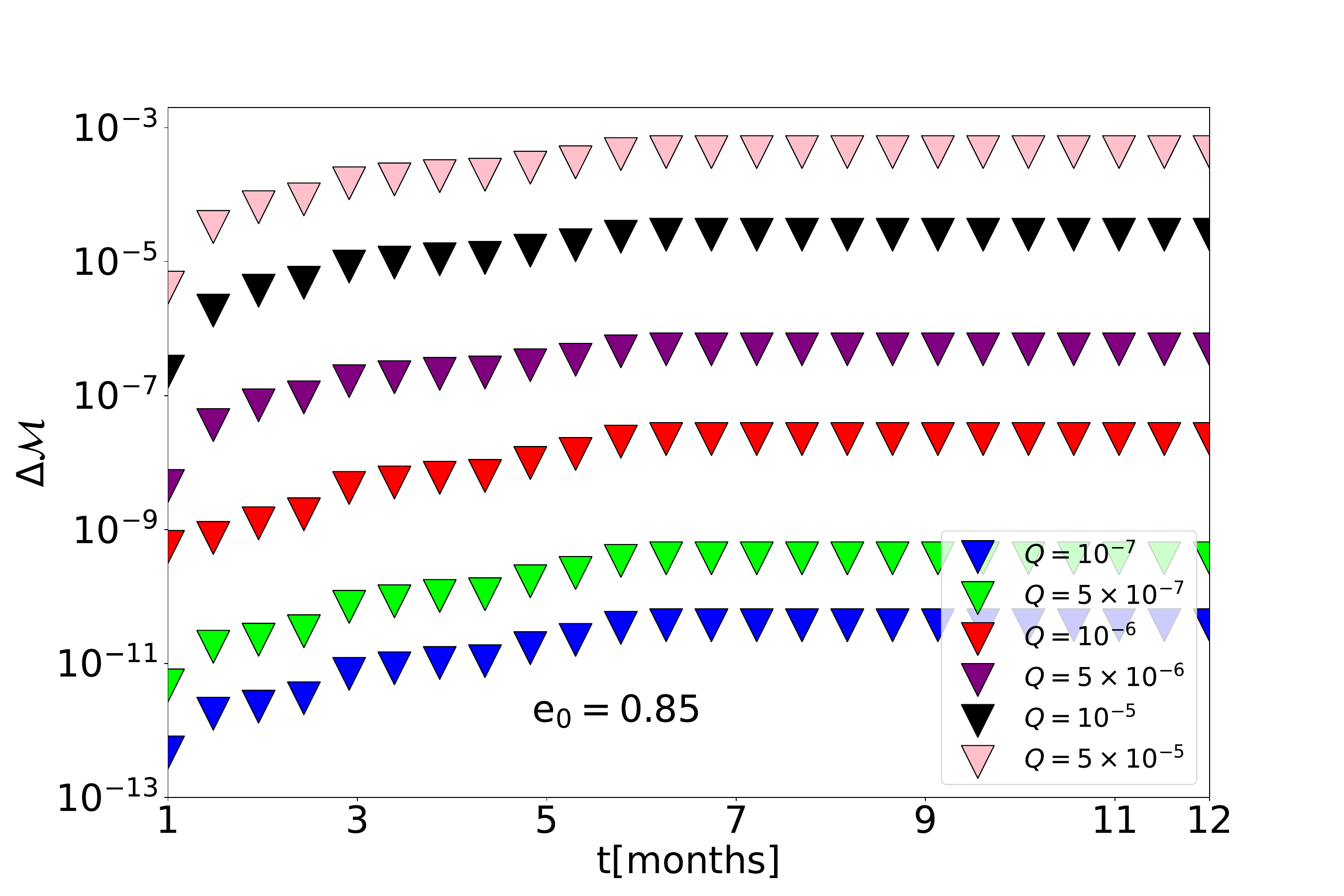}
\caption{Difference of mismatch as a function of observation time for four initial orbital eccentricitiy $e_0\in\{0.1,0.4,0.85\}$ is plotted, the other parameters are same with Fig.~\ref{Fig:mismatch1}.} \label{Fig:mismatch3}
\end{figure*}
Furthermore, we attempt to examine the difference between the mismatch ($\Delta\mathcal{M} = \vert\mathcal{M}^{\textup{MTE}}-\mathcal{M}^{\textup{DF-TEU}}\vert$) obtained from the DF-Teukolsky and MTE approaches. This analysis implies the order of magnitude at which the difference starts appearing in the mismatch for distinct values of ($e_0, Q$). 
\begin{alphalist}
\item In Fig. (\ref{Fig:mismatch3}) with $e_0=(0.1, 0.4, 0.85)$, we determine that the maximum change in the mismatch ($\Delta\mathcal{M}$) between these two formalisms goes from $\sim 10^{-4}$ to $\sim 10^{-3}$ (from the pink plots for $Q=5\times 10^{-5}$).
This growing deviation with eccentricity, as shown in Fig. (\ref{Fig:mismatch3}), suggests that highly eccentric EMRIs are more relevant to non-GR aspects, while the DF-approximation is sufficient to get an estimate for the bound on $Q$, but the MTE framework offers a more sensitive testbed for detecting non-GR effects.
\end{alphalist}

It should be noted that our analysis focuses on small values of $Q$, which is also consistent from a perturbative perspective and further supports the findings of previous studies based on DF approximation \cite{Rahman:2022fay, Zi:2024dpi, Kumar:2025njz}. Importantly, the correction of the tidal charge from the MTE results is part of the theory contribution that arises from the non-Ricci-flatness of the spacetime; however, a full and more accurate description necessitates the incorporation of additional technical components, such as the consistent inclusion of the electric Weyl tensor, within the MTE formalism. This includes both a more rigorous treatment of the underlying geometry and enhanced computational techniques to generate accurate and efficient waveform data\textemdash an essential aspect that will be pursued in future work. Similar studies within the post-Newtonian (PN) framework have also been carried out in \cite{Kumar:2024utz, AbhishekChowdhuri:2023gvu, Carson:2020dez}, where various parameters can be mapped to distinct non-GR theories, highlighting the need to encode the underlying theory in observable quantities. Nonetheless, the results presented here serve as an estimate of the detectability of tidal charge with a contribution from the non-Ricci-flatness of the metric that effectively appears in the MTE potential. 
They highlight its relevance for future GW observations while emphasizing the need for further refinement to enhance the precision and robustness of waveform predictions in such extended frameworks. Moreover, our results from the DF and MTE frameworks are equally applicable to the RN spacetime, assuming the influence of additional fields is ignored. Extending the MTE implementation to the RN case remains challenging due to the inherent coupling between gravitational and electromagnetic perturbations.

\section{Discussion}\label{dscn}
EMRIs are distinctive binary systems that have drawn considerable attention in GW astronomy due to their exceptional potential to probe the strong-field regime of gravity and test alternative theories beyond GR, particularly through future space-based detectors. Within this context, the possibility of detecting signatures of extra dimensions via GW observations has been an active area of research. Previous studies, such as \cite{Rahman:2022fay, Zi:2024dpi, Kumar:2025njz}, have explored EMRI dynamics assuming circular and eccentric orbits, typically employing the DF-Teukolsky equation that carries the same mathematical form as originally developed within the Ricci-flat vacuum GR framework. In other words, within the \textit{DF approximation}, we neglect the perturbation of the projected electric Weyl tensor, i.e., $\delta E_{\mu\nu}=0$, motivated by the assumption that the bulk backreaction effects on the brane are negligible, a valid low-energy approximation \cite{Kanno:2002ia} which  can also be equivalently understood as a small-charge approximation from the perspective of 4D effective metric (braneworld metric) on the brane about which we are ultimately studying the perturbations, where the influence of the extra dimension enters only through the static tidal charge, sitting in the metric function $\Delta (r)$. Since the bound originates from a leading-order approximation, its magnitude is expected to remain largely unchanged, although the observables will gain precision as additional physics coming from $\delta E_{\mu\nu}$ contributions is included. This article explores this aspect within the MTE framework. 
In this paper, we adopt the MTE framework for a $tr$-symmetric spacetime, which incorporates corrections to the Teukolsky potential arising from non-Ricci-flatness of the braneworld geometry, by relaxing the DF approximation. In such a symmetry, the NP tetrad and scalars simplify, many Ricci source terms vanish, gauge freedom can be effectively applied, and decoupled perturbation equations survive. Motivated by the systematic perturbative approach of \cite{Li:2022pcy, Guo:2023wtx, Guo:2024bqe}, our setup not only allows a robust cross-check of our previous results but also provides a next step for extending the analysis beyond GR, which will pave the way for more accurate and precise modeling of EMRI waveforms in non-GR spacetimes.

Several interesting things emerge from our present study, which we list below.
\begin{alphalist}
    \item The central objective of this study is to perform a detailed comparison between the results obtained using the MTE and those derived from the DF-Teukolsky formalism with modified $\Delta (r)\,.$ To this end, we compute key waveform diagnostics\textemdash specifically, the GW dephasing and mismatch\textemdash for various combinations of eccentricity $e_0$ and tidal charge $Q$, using both the DF-Teukolsky and MTE formalisms. Our findings reveal that, \textit{the order of magnitude of the constraint on $Q$ coming from imposing the detectability criteria remains largely the same for both approaches, indicating the consistency with previous DF-based studies to get an order of magnitude estimate of the bound on tidal charge} \cite{Rahman:2022fay, Zi:2024dpi, Kumar:2025njz}. Hence, DF-approximation, which is easier to implement in general than the MTE framework, which becomes more involved as one tries to incorporate more and more corrections from beyond GR theories, can be used with good effectiveness to obtain an estimate for the bound on the beyond GR parameter after imposing the detection threshold criteria as discussed before.
    \item Beside the above observation, we also find that the mismatch (Eq. (\ref{mismatch})) between the waveforms for $ Q\neq 0$ and $Q=0$ is consistently larger when computed using the MTE compared to DF-Teukolsky by incorporating the \textit{effect of non-Ricci flatness}, indicating a greater sensitivity to non-GR corrections, especially for higher eccentricity as shown in Fig.~(\ref{Fig:mismatch3}). This shows the importance of MTE based studies for \textit{precision measurement of GW observables}. While the DF approximation is effective in getting an order of magnitude estimate for the bound on the tidal charge, the MTE approach offers us a tool for performing a more precise study of the GW observable by incorporating the effect of an additional contribution coming from the non-vanishing variation of the projected Weyl tensor on the brane. This shows the importance of this study presented in this paper.
    \item Last but not least, our findings reveal that both the GW dephasing and the mismatch coming from both the DF-Teukolsky and MTE approach become larger as we increase the eccentricity. In particular, with the mismatch analysis, we find that for an eccentric orbit with $e_0=0.1$, the tidal charge can be constrained to an order of magnitude of $\sim 10^{-5}$, while for a more eccentric orbit with $e_0=0.85$, the constraint becomes tighter, improving to $\sim 5\times 10^{-6}$. This trend suggests that higher eccentricity enhances the detectability of extra-dimensional effects, making EMRIs with larger $e_0$ especially valuable for probing beyond GR physics. A similar behavior is observed in the dephasing analysis, which for both formalisms grows with increasing eccentricity.
\end{alphalist}
While the present work incorporates part of the theory correction emerging from the non-Ricci-flatness of the spacetime, encoded in the MTE potential, it highlights the need for a more comprehensive treatment that includes the full contributions from the underlying gravity theory. As in this work, we restricted attention to the spherically symmetric ($tr$-symmetric) case, which is Petrov type D and not Ricci-flat. Next, by contrast, when $tr$-symmetry is not maintained with black holes beyond GR, the spacetimes have new NP tetrads, non-vanishing Weyl scalars (e.g., $\Phi_{00},\Phi_{22}$), and perturbed directional derivatives appear, limiting the previous gauge-fixing technique and preventing straightforward decoupling. Our present results thus provide a concrete first step toward probing GW signatures in braneworld black holes and motivate extensions to rotating and spherical non $tr$-symmetric backgrounds ($g_{tt}g_{rr}\neq -1$), which we aim to focus in future works.

Further, the case of RN is considerably more involved than the braneworld background. Although the two metrics appear similar, RN is a solution of the Einstein-Maxwell equations, where the background curvature is supported by the electromagnetic field. Consequently, perturbations of RN necessarily couple gravitational and electromagnetic degrees of freedom, leading to a system of coupled equations rather than a single Teukolsky or master perturbation equation \cite{doi:10.1142/S0218271825400036, Giorgi:2023fqi} as in the Kerr case or in our present analysis. Previous works have explored this coupled structure, including the early treatment by Dudley and Finley, who studied perturbations under the approximation of fixing either the geometry or the electromagnetic field \cite{Dudley:1977zz, Dudley:1978vd}. Later studies showed that the Dudley-Finley approach provides a reasonable approximation only for a small charge \cite{Berti:2005eb} which is equivalent to $\delta E_{\mu\nu}=0$ assumption considered in our existing works \cite{Rahman:2022fay, Zi:2024dpi, Kumar:2025njz}. A more detailed treatment using axial/polar perturbations and consistency checks with the DF approximation can be found in \cite{Berti:2005eb,Saha:2025nsg}. In contrast, the braneworld geometry carries an effective charge without any Maxwell field on the brane, making the analysis considerably more tractable. For the RN case, using the DF approximation would yield a similar threshold value of the charge (positive only), consistent with our current findings; however, a more complete treatment should account for the contributions from the electromagnetic field and their coupling with spacetime curvature. We next aim to extend the MTE framework to the RN spacetime as well as braneworld scenario (based on two parameter expansion scheme \cite{Li:2022pcy} that includes perturbative combination of GW strength and beyond GR correction) with a unified treatment of coupled gravitational-electromagnetic perturbations and provide deeper insights into EMRI dynamics in charged or non-vacuum backgrounds \cite{Maartens:2010ar, Whisker:2008kk}. Moreover, in light of the present investigation, the inclusion of additional (perturbative) corrections of the field is expected to improve the overall precision of waveform modeling, as such corrections are perturbative in nature and small in order of magnitudes, while leaving the constraints on the corresponding field parameter (e.g. arising from the braneworld or RN background) nearly unaffected. Such an extension would mark a significant step toward testing gravitational theories beyond GR with higher precision and broader applicability.


To end, although our present analysis remains an approximation (incorporating non-Ricci-flat nature of the background), it represents an important initial step in this direction. The results clearly illustrate that even the non-Ricci-flatness contribution, when incorporated through the MTE framework, can lead to an increase in the precision while computing GW observables. This highlights the importance of developing refined perturbative tools capable of accurately reflecting the structure of alternative gravitational theories in the context of \textit{precision} GW astrophysics.

\section*{Acknowledgements}
The research of S.K. is funded by the National Post-Doctoral Fellowship (N-PDF: PDF/2023/000369) from the ANRF (formerly SERB), Department of Science and Technology (DST), Government of India. T.Z. is funded by the China Postdoctoral Science Foundation with Grant No. 2023M731137 and the National Natural Science Foundation of China with Grant No. 12347140 and No. 12405059. AB is supported by the Core Research Grant (CRG/2023/005112) by the Department of Science and Technology Science and Anusandhan National Research Foundation (formerly SERB), Government of India. A.B. also acknowledges the associateship program of the Indian Academy of Science, Bengaluru. Authors thanks Rong-Zhen Guo and Pratik Wagle for useful correspondence.

\appendix
\section{Source}\label{source}
The source term in the Teukolsky perturbation equation (in the background of (\ref{metric})) is written as
\begin{equation}\label{source_term}
\mathcal {T} _ {\ell m \omega} =  4 \int dt d\theta\sin\theta d\phi \frac{\left(B' _ 2 + {B' _ 2}^*\right)}{\bar{\rho}\rho^5} S_{\ell m\omega} e^{-  i (m\phi+ \omega t)},
\end{equation}
where,
\begin{align}
B' _ 2 =& - \frac{1}{2} \rho^8\bar {\rho}\mathcal {L} _{-1}
\bigg[\frac{1}{\rho^4}\mathcal{L}_0\bigg[\frac{T_{nn}}{\rho^2\bar{\rho}} \bigg]\bigg]
 \\ &-\frac{1}{2\sqrt{2}}\Delta^2 \rho^8\bar{\rho}\mathcal{L}_ {-1}\bigg[\frac{\bar{\rho}^2}{\rho^4}
J_+\bigg[\frac{T_{\overline{m}n}}{ \hat{\Delta} \rho^2\bar{\rho}^2} \bigg]\bigg] \ , \\
 {B' _ 2}^*=& - \frac {1} {4}\Delta^2 \rho^8\bar{\rho} J_+\bigg[\frac{1}{\rho^4}J_+
\bigg[\frac{\bar{\rho}}{\rho^2}T_{\overline{m}\overline{m}}\bigg] \bigg]
\\ & - \frac{1}{2\sqrt {2}}\Delta^2 \rho^8\bar{\rho} J_+ \bigg[\frac{\bar{\rho}^2}{\Delta \rho^4}\mathcal {L}_
{-1}\bigg[\frac{ T_ {\overline{m}n}}{\rho^2\bar {\rho}^2}\bigg] \bigg] \ ,
\end{align}
where we remind that $\Delta = r^{2}-2Mr+QM^{2}$, $K=r^{2}\omega$, $\rho = 1/r = \bar{\rho}$ and operators.
\begin{equation}
\begin{aligned}
J_+ =& \frac{\partial}{\partial r}+\frac{i K}{\Delta} \hspace{0.1cm} ; \hspace{0.1cm}\mathcal{L} _s = \frac{\partial}{\partial\theta}+\frac{m}{\sin \theta}+s \cot\theta\,, \\
\mathcal{L} _s^\dagger =& \frac{\partial}{\partial\theta}-\frac{m}{\sin \theta} + s
\cot\theta\,.
\end{aligned}
\end{equation}
Quantities ($T_{nn}, T_{\bar{m}n}, T_{\bar{m}\bar{m}}, T_{\bar{m}n}$) represent the projections of the stress-energy tensor onto the NP tetrad basis. Following \cite{Sasaki:2003xr, PhysRevD.102.024041, Zi:2024dpi}, we can write down the stress-energy tensor of the point particle (mass $\hat{m}$) in the following way,
\begin{align}\label{source1}
T^{\mu\nu} &= \frac{m_{\textup{SO}} (dt/d\tau)^{-1}}{\Sigma \sin\theta}\frac{dz^{\mu}}{d\tau}\frac{dz^{\nu}}{d\tau}
\nonumber\\
&\times\delta\left[r-r(t)\right]\delta\left[\theta -\theta(t)\right]\delta\left[\phi -\phi(t)\right]
\end{align}
where $z^{\mu} = \left[t, r(t), \theta(t), \phi(t)\right]$ is the geodesic trajectory with proper time $\tau = \tau(t)$. Eq. (\ref{source1}) uses geodesic velocities, which can be replaced using Eq. (\ref{geodesic}). Thus, tetrad components of the stress-energy tensor are given by:

\begin{equation}
\begin{aligned}
T_{nn} =& \frac{m_{\textup{SO}} C_{nn}}{\sin\theta}
\delta\left[r-r(t)\right]\delta\left[\theta -\theta(t)\right]\delta\left[\phi -\phi(t)\right]\,, \\
T_{\bar{m}n} =& \frac{m_{\textup{SO}} C_{\bar{m}n}}{\sin\theta}
\delta\left[r-r(t)\right]\delta\left[\theta -\theta(t)\right]\delta\left[\phi -\phi(t)\right]\,, \\
T_{\bar{m}\bar{m}} =& \frac{m_{\textup{SO}} C_{\bar{m}\bar{m}}}{\sin\theta}
\delta\left[r-r(t)\right]\delta\left[\theta -\theta(t)\right]\delta\left[\phi -\phi(t)\right]\;,
\end{aligned}
\end{equation}
where,
\begin{equation}
\begin{aligned}
C_{nn} =& \frac{1}{4\Sigma^{3}}\Big(\frac{dt}{d\tau}\Big)^{-1} \Big(Er^{2}+\Sigma\frac{dr}{d\tau} \Big)^{2}\,, \\
C_{\bar{m}n} =& -\frac{\rho}{2\sqrt{2}\Sigma^{2}}\Big(\frac{dt}{d\tau}\Big)^{-1} \Big(Er^{2}+\Sigma\frac{dr}{d\tau} \Big)\Big(-iL_{z}\csc\theta+\Sigma\frac{d\theta}{d\tau}\Big)\,, \\
C_{\bar{m}\bar{m}} =& \frac{\rho^{2}}{2\Sigma}\Big(\frac{dt}{d\tau}\Big)^{-1}\Big(-iL_{z}\csc\theta+\Sigma\frac{d\theta}{d\tau} \Big)^{2}\,.
\end{aligned}
\end{equation}
As mentioned earlier, we use Eq. (\ref{geodesic}) to replace geodesic velocities. Also, note that we replace the parameterization in Eq.~\eqref{parametrize} to perform the analysis for eccentric orbits. Thus, it constructs the source term that we use to compute GW fluxes in section (\ref{perturbation}).


\end{document}